\documentclass{article}
\usepackage{amssymb}
\usepackage[centertags]{amsmath}

\begin{document}

\title{The renormalization  method based on the Taylor expansion and applications for asymptotic analysis}
\author{ Cheng-shi Liu \\Department of Mathematics\\Northeast Petroleum University\\Daqing 163318, China
\\Email: chengshiliu-68@126.com}

 \maketitle

\begin{abstract}

 Based on the Taylor expansion, we propose a renormalization method for asymptotic
 analysis. The standard renormalization group (RG) method for asymptotic analysis can be derived out from this new
 method, and hence the mathematical essence of the RG method is also recovered.
The biggest advantage of the proposed method is that
 the secular terms in perturbation series are automatically eliminated, but in usual
 perturbation theory, we need more efforts and tricks to eliminate
 these terms.  At the same time,  the mathematical foundation of
 the method is simple and the logic of the method
  is very clear, therefore, it is very easy in practice. As
 application, we obtain the uniform valid asymptotic solutions to
 some problems including vector field,  boundary layer and boundary value problems of nonlinear wave
 equations. Moreover, we discuss the normal form theory and
 reduction equations of dynamical systems.
 Furthermore, by combining the topological deformation
  and the RG method, a modified method namely
 the homotopy renormalization  method (for simplicity, HTR) was
proposed to overcome the weaknesses of the standard RG
 method. In this HTR method, since  there is a freedom to choose the first order
approximate solution in perturbation expansion, we can improve the
global solution. In particular, for those equations including no a
small parameter, the HTR  method can also be applied. Some concrete
applications including multi-solutions problems, the forced Duffing
equation and the Blasius equation are given.


\end{abstract}

\section{Introduction}
Renormalization method has become a famous method, and is perhaps
one of the most important methods in physics. In 1948, Schwinger,
Feynman and Tomonaga [1-6] introduced independently the
renormalization method to remove the divergences in the perturbation
theory arising in the quantum electrodynamics. Dyson proved three
methods were equivalent[7,8]. For the extremely high precision,
quantum electrodynamics has become the most accurate theory.  In
1953, Gell-mann and Low [9] proposed the renormalization group (for
simplicity, RG) method to handel the divergences in quantum field
theory. In 1971, Wilson[10] introduced the RG method to develop the
theory of the second order phase transitions and hence brought the
statistical physics into a new era.

In 1990s, Goldenfeld et al initially used the RG method instead of
dimensional analysis, to nonlinear partial differential
equations[11-20]. By considering the Cauchy data as the analogues of
the bare quantities in quantum field theory which can be
renormalized by perturbation, a clear physical interpretation to the
RG method was given.  Their results showed that the RG method was
more efficient and perhaps more accurate than the usual perturbation
methods for asymptotic analysis because it
 avoided the necessity to perform asymptotic matching. Then the RG
 method were developed and applied from both of theory and
 applications. For example, Kunihiro, Nozaki, Oono and Shiwa proposed the proto-RG method in[21-23],
  and Tu and Cheng introduced an improved renormalization group
  method to perturbed PDEs[24-27]. Other applications can be found
  in a large number of references.

 Although the RG method unifies
some perturbation methods including singular and reductive
perturbation theory and hence shows its advantages, such as no need
to guess unexpected fractional power laws or logarithmic functions
of small parameter $\epsilon$, and seems also no need to perform
asymptotic matching, the mathematical foundation of the RG method
for asymptotic analysis is still not clear. Kunihiro[28-36] tried to
give the RG method a geometrical interpretation  based on the
classical theory of envelop in differential geometry. Furthermore
the standard RG method and its geometrical formulas have been
developed and applied to many problems [37-56] such as center
manifolds[39], quantum kinetics[41,47], normal form theory[42],
invariant manifold and reduction equation[43-45,52], variation of
parameters[48] and resonance of quantum[51] and so forth. In
addition, the error estimate of approximate solutions obtained by
the RG method are given in [37,39,42], and comparison of RG method
and homotopy analysis is discussed in [55]. However, both of the
standard RG method and its geometrical interpretation are still not
satisfied. For example, in the standard RG method, we need to
introduce some auxiliary variables and some renormalized auxiliary
parameters and eliminate them by the RG equation, this is a complex
process. In the geometrical formulation, we need a key assumption on
the point of tangency to simplify the RG equation, this is also not
a natural treatment. For both of RG method and conventional
asymptotic methods, at a purely technical level, the starting point
is the removal of divergences from a perturbation series[17]. Thus,
if there is a method for which it is no need to consider the secular
terms, this method will be interesting and will provide some
insights on the essence of the perturbation problems. In the present
paper, I will give such a renormalization method.

  Our renormalization method is only based simple Taylor
 expansion, by which the renormalization group method for asymptotic
analysis can be derived out, in other words, the RG method is just
based on the usual Taylor expansion, and all key assumptions in the
standard RG method can be given as the natural facts in Taylor
expansion. The mathematical foundation of the proposed
renormalization method(TR, for simplicity) is so simple and clear
that we can easily understand it in theory and use it in practice.
As applications, we use the proposed renormalization method to deal
with a lot of singular and regular perturbation problems and obtain
their uniform valid asymptotic solutions and other behaviors such as
normal form theory and reduction equations. The biggest advantage of
our method is that the secular terms can be automatically
eliminated. On the other hand, the proposed renormalization method
may be more simple and direct than other methods, since it does not
require the asymptotic matching and not need the introduction of
auxiliary renormalization parameters.

 However, for the RG methods, there also exist some weaknesses(for concrete examples,
 see section 5.1), for example,
  how to
choose the equations determining integral constants needs a clear
guidance, moreover, for some equations, the standard RG method can
not work or do not improve the global solutions. In final, if there
does not exist a small parameter in the equations, the RG or TR
method can not be applied. In order to overcome these weaknesses,
based the TR method, we propose a deformed renormalization method
namely homotopy renormalization method (HTR for simplicity) to solve
these problems. Some applications are given by the HTR method.

This paper is organized as follows. In section 2, by an example, we
describe the standard renormalization group method for asymptotic
analysis and its geometrical interpretation. In section 3, we
propose a renormalization method based on the Taylor expansion and
give a detailed discussion on the mathematical foundation of the
method.  In section 4, we give some typical applications of the
renormalization method, including a vector filed problem, two
nontrivial boundary layer problems which are the singular
perturbation problems,  the boundary value problems of two nonlinear
wave equations arising from mechanics of continuous medium, and
normal form theory and reduction equations of dynamical systems. In
section 5, we discuss the weaknesses of the RG method by some
examples, and give the homotopy renormalization method (HTR) by
combining a topological deformation with the TR method to overcome
these weaknesses. Furthermore,  some applications are given by HTR
method. The last section is a short conclusion.

\section{The standard RG method and its geometrical formula}
 By the following example, we describe
 the standard renormalization group method and the Kunihiro's geometrical
interpretation. Consider the Rayleigh equation
\begin{equation}
\ddot{y}+y=\varepsilon(\dot{y}-\frac{1}{3}\dot{y}^3),
\end{equation}
where $\varepsilon$ is a small parameter. The standard
renormaoization group method [11] solving this equation is as
follows. First, expanding $y$ as $y=y_0+\varepsilon y_1+...$ gives
\begin{equation}
y(t,
t_0)=R_0\sin(t+\theta_0)+\varepsilon[(\frac{R_0}{2}-\frac{R_0^3}{8})(t-t_0)\sin(t+\theta_0)+
\frac{R_0^3}{96}\cos3(t+\theta_0)]+O(\varepsilon^2).
\end{equation}
Second, by introducing  an arbitrary time $\tau$,  the term $t-t_0$
is splitted as $t-\tau+\tau-t_0$ and  the terms containing
$\tau-t_0$ are absorbed into the renormalized counterparts $R$ and
$\theta$ of $R_0$ and $\theta_0$. And then introduce two
renormalization constants $Z_1=1+\sum_{1}^{\infty}a_n\varepsilon^n$
and $Z_2=1+\sum_{1}^{\infty}b_n\varepsilon^n$ such that
$R_0(t_0)=Z_1(t_0, \tau)R(\tau)$ and
 $\theta_0(t_0)=Z_2(t_0, \tau)+\theta(\tau)$, where $R$ and $\theta$ are two functions of $\tau$.
 Since $\tau$ does not appear in the original equation, so $\frac{\partial y}{\partial \tau}=0$
which is the RG equation. By choosing special coefficients $a_n$ and
$b_n$ to eliminate terms containing $\tau-t_0$, the global
approximate solution will be given.

Next, in the Kunihiro's geometrical interpretation, the solution (2)
is considered as a family curves parameterized by $t_0$, and
 $R_0$ and $\theta_0$ are functionally dependent on $t_0$,
 Then the global solution is the envelop of the curves
(2) by eliminating $t_0$ from the equation (2) and the following
equation
\begin{equation}
\frac{\partial y(t, t_0)}{\partial t_0}=0.
\end{equation}
The last equation is just the RG equation. But the  RG equation is
in general  a complicated equation involving $R_0(t_0)$ and
$\theta_0(t_0)$ and their derivatives, so Kunihiro assumes that the
parameter $t_0$ coincides with the point of tangency, that is,
$t=t_0$, by which the RG equation can be actually great simplified.
According to Eq.(2) and Eq.(3), we obtain

\begin{equation}
\dot{R_0}(t)=\varepsilon(\frac{R_0(t)}{2}-\frac{R_0^3(t)}{8}),
\dot{\theta}=0 \end{equation}
  whose solutions are
\begin{equation}
R_0(t)=\frac{R_0(0)}{\sqrt{e^{-\varepsilon t}+(1-e^{-\varepsilon
t})/4}},\theta_0(t)=\theta_0,
\end{equation}
where $\theta_0$ is a constant. Therefore, the corresponding envelop
is given by
\begin{equation}
y_E(t)=y(t,
t)=R_0\sin(t+\theta_0)+\varepsilon\frac{R_0(t)^3}{96}\cos3(t+\theta_0),
\end{equation}
which is a global approximate solution to Eq.(1).

 For the standard RG
method, it is relatively complicated in mathematical treatment since
we need to split the secular terms and introduce renormalization
constants to adjust the bare parameters. At the same time, for the
geometrical interpretation of the RG method,  the condition $t_0=t$
is an assumption but not a natural fact.  In next section, we will
show that based on the simple Taylor expansion theory, the RG method
and its geometrical formula can be easily obtained. Among these, the
RG equation can be derived out naturally and can be replaced by a
basic fact in Taylor expansion.

\section{The renormalization method based on the Taylor expansion}

Based on the envelop theory in differential geometry, Kunihiro gave
the RG method a geometrical interpretation. By considering the
perturbation solution as a local curve parameterized by $t_0$, the
global solution can be obtained by taking the envelop of all local
solutions. This treatment gives the same results with that by the RG
method, and this is a beautiful interpretation for the RG method.
However, the RG equation is still an assumption and another key
assumption of taking $t_0$ as the tangent point is also needed. We
believe that behind the RG method has yet some unknown structures
which need to be recovered. It is surprising for me that this
structure is just the usual Taylor expansion.

 For a given
differential equation with initial conditions, we can use the Taylor
series expansion at the initial point to find its local solution. We
have the Cauchy theorem for ordinary differential equation, which
says that if $F(x, t)$ is an analytic function in $|t-t_0| <\alpha,
|x - x_0|<\beta$, then there exists
 unique analytic solution in the neighborhood of $t_0$ to the equation $\frac{dx}{dt}=F(x,t)$ with original
 condition $x(t_0) = x_0$. For partial differential equation, we have the Cauchy-Kowalevskaya theorem.
 In general, to enlarge the convergence region to give the global
 solution,
 a standard and classical method is the analytic continuation.  For
 the constructions of the global solutions to nonlinear differential
 equations, the analytic continuation method is essential even
 thought it is so complicated. Some methods of constructing global solutions, such as
the homotopy analysis method[57] and the general series
method[58-61], all dependent on the analytic continuation
essentially from the computation point of view. However, the general
power series expansion method gives us some new insights.  In fact,
the way of the general power series method is to expand the solution
at a general point $t_0$ as follows
\begin{equation}
y(t)=\sum_{n=0}^{+\infty}b_n(t_0)(t-t_0)^n.
\end{equation}
If $t_0$ is a general point and is considered as a parameter, we
arrive at the key point, that is, $b(t_0)$ is exactly the solution!
Another key relation is $b'_{n-1}=nb_n$ for $n=1,2,\cdots$. In
particular,  the equation $b'_0=b_1$  will replace the RG equation
in the RG method. By these two simple facts, all things in the RG
method can be clearly and simply explained and proved. In fact,  if
we take the first approximation $y_1(t)=y(t_0)+y'(t_0)(t-t_0)$ which
is the tangent line of the solution curve at the tangent point
$t_0$,
 the envelop of these tangent lines is just the solution. If we
take other finite order approximation, their envelop is also the
solution. Although $y(t_0)$ is just the exact solution, we have no
general method to solve it except the analytic continuation.
Therefore,  to find the local approximate solution is a key step.  A
possible method is to use perturbation technics to expand the
solution as a power series of some small parameter in the equation
and take the approximate solution with local parameter $t_0$ as the
local solutions. And then, by the RG method, taking the envelop of
these local solution  gives the global solution.  This is just the
basic idea of the construction of the global solution by the RG
method  based the envelop theory and the geometrical meaning of the
global solution.

Now we give a new renormalization method and hence also provide a
strict mathematical foundation of the RG method and its geometrical
interpretation. Consider a differential equation
\begin{equation}
N(y)=\epsilon M(y),
\end{equation}
where $N$ and $M$ are in general linear or nonlinear operators.
Assuming that the solution can be expanded as a power series of the
small parameter $\epsilon$
\begin{equation}
y=y_0+y_1\epsilon+\cdots+y_n\epsilon+\cdots,
\end{equation}
and substituting it into the above equation yields the equations of
$y_n$'s such as
\begin{equation}
N(y_0)=0,
\end{equation}
and
\begin{equation}
N_1(y_1)=M_1(y_0),\cdots,N_k(y_k)=M_k(y_{k-1}),\cdots
\end{equation}
for some operators $N_k$ and $M_k$. By the first equation, we give
the general solution of $y_0$ including some integral constants $A$
and $B$ and so on. In general the number of the integral constants
is equal to the order of the differential equation. Then we find the
particular solutions of $y_k$ which sometimes include also some
integral constants, and expand them as the power series at a general
point $t_0$ as follows,
\begin{equation}
y_k(t)=\sum_{m=0}^{+\infty}y_{km}(t_0)(t-t_0)^m, k=0,1,\cdots.
\end{equation}
Therefore, by rearranging the summation of these series, we obtain
the final solution
\begin{equation}
y(t)=\sum_{n=0}^{+\infty}Y_n(t_0,\epsilon)(t-t_0)^n,
\end{equation}
where
\begin{equation}
Y_n(t_0,\epsilon)=\sum_{k=0}^{+\infty}y_{kn}(t_0)\epsilon^k,
n=0,1,\cdots.
\end{equation}
The formula (13) is the most basic formula for our theory from which
we can give every thing of the standard RG method and more.
Concretely, we have

(i). $y(t)=Y_0(t,\epsilon)$;

(ii).$Y_n(t,\epsilon)=\frac{1}{n!}\frac{\mathrm{d}^{n}}{\mathrm{d}t^n}Y_{0}(t,\epsilon),n=1,2,\cdots$;

(iii). $\frac{\partial^n}{\partial t_0^n}y(t,t_0)=0, n=1,2,\cdots$.

These results are the simple facts in the Taylor expansion but it is
the most important result for our theory by the following reasons.
In fact, formula (i) tells us that the solution is just given by the
first term $Y_0$ of the expansion when we consider $t_0$ as a
parameter, and hence all other terms need not be considered at all!
However, in general, this first term includes some integral
constants which need to be determined. The so-called renormalization
is just to do this thing. The point in our theory is to take $m$
relations in case (ii) as the equations to determinate the $m$
unknown parameters rather than to use the initial conditions to do
it. The third formula (iii) gives us automatically the RG equation
in the standard RG method
\begin{equation}
\frac{\partial}{\partial t_0}y(t,t_0)=0,
\end{equation}
and the assumption $t=t_0$ in its geometrical explanation can also
be derived out naturally.

\textbf{Definition}. We call the relations
\begin{equation}
Y_n(t,\epsilon)=\frac{1}{n!}\frac{\mathrm{d}^{n}}{\mathrm{d}t^n}Y_{0}(t,\epsilon),n=1,2,\cdots
\end{equation}
as the renormalization equations.

Now according to the formulas, we can give the key theoretical
foundation and main steps of new renormalization method as follows:

Firstly, $Y_0(t,\epsilon)$ is just exact solution by expansion (13),
in which there are some constants to be determined.

Secondly, the natural relation $Y_0'=Y_1$ (if there are $m$ unknown
constants, we will take $m$ renormalization equations
$Y_n(t,\epsilon)=\frac{1}{n!}\frac{\mathrm{d}^{n}}{\mathrm{d}t^n}Y_{0}(t,\epsilon),n=1,2,\cdots,m$)
give the renomalization equations satisfied by unknown constants,
which can be used to determine these unknown constants in $Y_0$.

Thirdly, by renomalization equations, solving out the unknown
constants and substituting them into $Y_0$ give the asymptotic
solution.

For simplicity, we will use \textbf{TR} method (i.e., Taylor
renormalization method) to call the new renormalization method based
on the Taylor expansion.

We must emphasize that if there are $m$ unknown integral constants,
we will need $m$ renormalization equations $Y'_{n-1}=nY_n$ to get a
closed equations system to solve out $m$ unknown functions for
$n=1,\cdots,m$. But, in practice, we usual need only one
renormalization equation, that is, the first renormalization
equation $Y'_{0}=Y_1$ to get the closed equations system by some
approximation or other balance relations.

\textbf{Remark 1}. For the system of finite or infinite-dimensional
ordinary differential equations, we only need to replace the
corresponding scale functions by the vector functions to give whole
theory and formulas.

Now we prove that the key assumptions in standard RG method and its
geometrical formula can be derived from our theory.  First, we prove
that the standard RG equation holds naturally, that is
\begin{equation}
\frac{\partial}{\partial t_0}y(t,t_0)=0.
\end{equation}
In fact, by direct computation, we have
\begin{equation}
\frac{\partial}{\partial
t_0}y(t,t_0)=\sum_{n=0}^{+\infty}Y'_n(t_0,\epsilon)(t-t_0)^n-
\sum_{n=1}^{+\infty}(n+1)Y_{n+1}(t_0,\epsilon)(t-t_0)^{n}.
\end{equation}
According to $Y'_{n-1}=nY_n=\frac{y^{(n-1)}}{(n-1)!}$, the
conclusion is right.

Then we derive out naturally the assumption $t_0=t$ in Kunihiro's
geometrical formulation. In fact, without loss of the generality, we
take the first approximation
\begin{equation}
y(t,t_0)=Y_0(t_0,\epsilon)+Y_1(t_0,\epsilon)(t-t_0)
\end{equation}
which is the tangent line of the solution curve at the point $t_0$.
So the envelop of these tangent lines is just the solution. If we
take other finite order approximation which is also tangential to
the solution curve, their envelop is also just the solution.
 Form Eq.(19), we have
\begin{equation}
\frac{\partial}{\partial
t_0}y(t,t_0)=Y'_0(t_0,\epsilon)+Y'_1(t_0,\epsilon)(t-t_0)-Y_1(t_0,\epsilon).
\end{equation}
According to  $\frac{\partial}{\partial t_0}y(t,t_0)=0$ and
$Y'_0=Y_1$, we must have $t_0=t$. The proof is completed.

By the above discussion, we have seen that the RG method and its
geometrical interpretation can clearly and simply be obtained from
the usual Taylor expansion.

For the purpose of comparison, we use TR method to solve again the
Rayleigh equation
\begin{equation}
\ddot{y}+y=\epsilon(\dot{y}-\frac{1}{3}\dot{y}^3),
\end{equation}
where $\epsilon$ is a small parameter. First, expanding $y$ as
$y=y_0+\varepsilon y_1+...$ gives
\begin{equation}
y(t,
t_0)=R_0\sin(t+\theta_0)+\epsilon[(\frac{R_0}{2}-\frac{R_0^3}{8})(t-t_0)\sin(t+\theta_0)+
\frac{R_0^3}{96}\cos3(t+\theta_0)]+O(\epsilon^2).
\end{equation}
The solution (22) is rewritten as
\begin{equation*}
y(t,
t_0)=R_0\sin(t_0+\theta_0)+\epsilon\frac{R_0^3}{96}\cos3(t_0+\theta_0)+
\{R_0\cos(t_0+\theta_0)
\end{equation*}
\begin{equation}
+\epsilon(\frac{R_0}{2}-\frac{R_0^3}{8})\sin(t+\theta_0)-
\epsilon\frac{R_0^3}{32}\sin3(t+\theta_0)\}(t-t_0)+O(\epsilon^2).
\end{equation}
The  renormalization equation is
\begin{equation*}
\frac{\partial}{\partial
t_0}(R_0\sin(t_0+\theta_0)+\epsilon\frac{R_0^3}{96}\cos3(t_0+\theta_0))=
R_0\cos(t_0+\theta_0)
\end{equation*}
\begin{equation}
+\epsilon(\frac{R_0}{2}-\frac{R_0^3}{8})\sin(t_0+\theta_0)-
\epsilon\frac{R_0^3}{32}\sin3(t_0+\theta_0).
\end{equation}

Here I must give an important remark. Since there are two unknown
functions $R(t_0)$ and $\theta(t_0)$ in this one renormalization
equation, in principle we need take the second approximation and
take another renormalization equation $Y'_1=2Y_2$. But we do not
intend to do this thing because the equations will become more
complicated even than the original equation and hence it is hard to
work. More important thing is that the renormalization equation is
only an approximate equation since the perturbation solution is
approximate itself. Therefore, we only need to deal with the
renormalization equation approximately and to find its solution as
exact as possible. At the same time, of course, we also need to make
the computation as simple as possible. Thus, in order to obtain a
simple and nontrivial closed equations about two unknown functions,
we must ignore some terms. For the purpose, we expand the
renormalization equation as
\begin{equation*}
R'_0\sin(t_0+\theta_0)+R'_0\cos(t_0+\theta_0)\theta'+\epsilon\frac{R_0^2}{32}R'_0\cos3(t_0+\theta_0)
-\epsilon\frac{R_0^3}{32}\sin3(t_0+\theta_0)\theta'
\end{equation*}
\begin{equation}
=\epsilon(\frac{R_0}{2}-\frac{R_0^3}{8})\sin(t_0+\theta_0)-
\epsilon\frac{R_0^3}{32}\sin3(t_0+\theta_0).
\end{equation}
It is easy to see that it is impossible to separate the functions
$R_0$ and $\theta_0$ unless we ignore some terms such as
$\epsilon\frac{R_0^2}{32}R'_0\cos3(t_0+\theta_0)$ and
$\epsilon\frac{R_0^3}{32}\sin3(t+\theta_0)$ under which we can get
two closed equations
\begin{equation}
 \theta'_0=0,
\end{equation}
\begin{equation}
R'_0=\epsilon(\frac{R_0}{2}-\frac{R_0^3}{8}).
\end{equation}
It is the same with the Goldenfeld's and Kunihiro's results, but the
TR method is more simple and direct.

The examples treated by standard RG method, including boundary layer
problems, turning point problems, reductive perturbation and
invariant manifolds, can be solved by the TR method. In order to
further illustrate our theory and method, we give other two
interesting examples as follows.

\textbf{Example 1}.  Consider the Mathieu equation
\begin{equation}
y''+(a+2\epsilon \cos t)y=0,
\end{equation}
where $a$ and $\epsilon$ are parameters. As done in paper [20],
taking $a=\frac{1}{4}+a_1\epsilon+\cdots$ and $y=y_0+\epsilon
y_1+\cdots$ gives
\begin{equation*}
y=R\cos(t/2+\theta)+\epsilon
R\{-\frac{1}{2}\cos(t/2+\theta)+\frac{1}{2}\cos(3t/2+\theta)
\end{equation*}
\begin{equation}
-a_1\sin(t/2+\theta)(t-t_0)-\sin(t/2-\theta)(t-t_0)\}+O(\epsilon^2).
\end{equation}
The renormalization equation is
\begin{equation*}
\frac{\partial}{\partial
t_0}(R\cos(t_0/2+\theta)-\frac{1}{2}\epsilon
R\cos(t_0/2+\theta)+\frac{1}{2}\epsilon
R\cos(3t_0/2+\theta))=-a_1\epsilon R\sin(t_0/2+\theta)
\end{equation*}
\begin{equation}
-\epsilon
R\sin(t_0/2-\theta)+\frac{1}{2}R\sin(t_0/2+\theta)+\frac{1}{4}\epsilon
R\sin(t_0/2+\theta)-\frac{3}{4}\epsilon R\sin(3t_0/2+\theta).
\end{equation}
A choice of the separation renormalization equations are given by
\begin{equation}
\theta=0,
\end{equation}
\begin{equation}
\{\cos(t_0/2)+\frac{\epsilon}{2}\cos(3t_0/2)-\frac{\epsilon}{2}\cos(t_0/2)\}R'=-(a_1+1)\epsilon
R\sin(t_0/2).
\end{equation}
Solving last equation yields
\begin{equation}
R(t_0)=R_0(\frac{\cos(t/2)}{1+\frac{2\epsilon}{1 -2\epsilon}
\cos(t/2)})^{\frac{2\epsilon}{1-2\epsilon})(1+a_1)}.
\end{equation}
The corresponding asymptotic solution is
\begin{equation}
y=R_0(\frac{\cos(t/2)}{1+\frac{2\epsilon}{1 -2\epsilon}
\cos(t/2)})^{\frac{2\epsilon}{1-2\epsilon})(1+a_1)}\{\cos(t/2)-
\frac{1}{2}\epsilon\cos(t/2)+\frac{1}{2}\epsilon\cos(3t/2)\}.
\end{equation}
This is a global uniform valid asymptotic solution when $a_1>-1$ and
$ \epsilon<\frac{1}{4}$. Moreover, this is a periodic solution.
Comparing with the RG method and other perturbation technics, the TR
method is more simple and direct.

\textbf{Example 2}. Consider the equation $y''+y=\epsilon y'$.

 In most examples, we use only one renormalization equation. In this example, we will use one renormalization equation
  and  two renormalization equations to solve it respectively.  By the standard procedure,
the second order perturbation solution is given by
\begin{equation}
y=R\sin(t_0+\theta)+(1+\frac{\epsilon}{2}-\frac{3\epsilon^2}{8})R\cos(t_0+\theta)(t-t_0)
+(-1-\frac{\epsilon}{2}+\frac{\epsilon^2}{2})R\sin(t_0+\theta)(t-t_0)^2+O(\epsilon^3).
\end{equation}

Firstly, if we take only the first one renormalization equation,  a
choice of closed separation equations is
\begin{equation}
R'=0,\theta'=1+\frac{\epsilon}{2}-\frac{3\epsilon^2}{8}.
\end{equation}
Solving it yields
\begin{equation}
R=R_0,\theta(t_0)=(1+\frac{\epsilon}{2}-\frac{3\epsilon^2}{8})t_0+\theta_0,
\end{equation}
and then the asymptotic solutions is
\begin{equation}
y=R_0\sin((2+\frac{\epsilon}{2}-\frac{3\epsilon^2}{8})t+\theta_0).
\end{equation}

Secondly, we take the first two renormalization equations
\begin{equation}
(R\sin(t_0+\theta))'=(1+\frac{\epsilon}{2}-\frac{3\epsilon^2}{8})R\cos(t_0+\theta),
\end{equation}
\begin{equation}
\{(1+\frac{\epsilon}{2}-\frac{3\epsilon^2}{8})R\cos(t_0+\theta)\}'=
2(-1-\frac{\epsilon}{2}+\frac{\epsilon^2}{2})R\sin(t_0+\theta).
\end{equation}
Solving it gives
\begin{equation}
R\sin(t+\theta)=A\cos(\sqrt{1-\epsilon-\epsilon^2}t)+B\sin(\sqrt{1-\epsilon-\epsilon^2}t),
\end{equation}
and hence the asymptotic solution is just
\begin{equation}
y=A\cos(\sqrt{1-\epsilon-\epsilon^2}t)+B\sin(\sqrt{1-\epsilon-\epsilon^2}t),
\end{equation}
where $A$ and $B$ can be determined by initial conditions.

We can see that these two asymptotic solutions are not the same one
but a little difference.

\section{Applications of the TR method}

\subsection{Application to vector field-Three wave regular
perturbation equations}

 In previous section, we take some scalar equations to solve by TR method. In this section, we deal with a vector
 field problem.  Here we use the TR
method to solve the asymptotic solution of three-wave equation[62]
\begin{equation}
x''+\omega^2_1x=-\epsilon yz,
\end{equation}
\begin{equation}
y''+\omega^2_2y=-\epsilon xz,
\end{equation}
\begin{equation}
z''+\omega^2_3z=-\epsilon xy,
\end{equation}
which is the simplest coupled system of three springs with the
second order interactions, where $\epsilon$ is a small parameter.
 Three-mode systems are interesting in theory and
application and hence are studied deeply[63-65].
Although the three-wave equation can be solved by some routine perturbation methods in
[66], the TR method is more simple. Expanding $x,y$ and $z$ as the
power series of $\epsilon$ such as $x=x_0+x_1\epsilon+\cdots$,
$y=y_0+y_1\epsilon+\cdots$, $z=z_0+z_1\epsilon+\cdots$, and
substituting them into the above equations give
\begin{equation}
x_0=A_1\cos(\omega_1t+\theta_1),
\end{equation}
\begin{equation}
y_0=A_2\cos(\omega_2t+\theta_2),
\end{equation}
\begin{equation}
z_0=A_3\cos(\omega_3t+\theta_3),
\end{equation}
and
\begin{equation}
x''_1+\omega^2_1x_1=-\frac{A_2A_3}{2}\{\cos((\omega_2+\omega_3)t+\theta_2+\theta_3)
+\cos((\omega_2-\omega_3)t+\theta_2-\theta_3)\},
\end{equation}
\begin{equation}
y''_1+\omega^2_2y_1=-\frac{A_1A_3}{2}\{\cos((\omega_1+\omega_3)t+\theta_1+\theta_3)
+\cos((\omega_1-\omega_3)t+\theta_1-\theta_3)\},
\end{equation}
\begin{equation}
z''_1+\omega^2_3z_1=-\frac{A_1A_2}{2}\{\cos((\omega_1+\omega_2)t+\theta_1+\theta_2)
+\cos((\omega_1-\omega_2)t+\theta_1-\theta_2)\}.
\end{equation}
There are the following three cases to be considered:

\emph{\textbf{Case (i)}}. $\omega_2+\omega_3\neq\omega_1$ and
$\omega_2-\omega_3\neq\pm\omega_1$. In the case it is obvious that
there exist no secular terms in these solutions, so the asymptotic
solutions are given by
\begin{equation*}
x=A_1\cos(\omega_1t+\theta_1)-\frac{A_2A_3\epsilon}{2}\{\frac{\cos((\omega_2+\omega_3)t+\theta_2+\theta_3)}{\omega^2_1-(\omega_2+\omega_3)^2}
+\frac{\cos((\omega_2-\omega_3)t+\theta_2-\theta_3)}{\omega^2_1-(\omega_2-\omega_3)^2}\}+O(\epsilon^2),
\end{equation*}
\begin{equation*}
y=A_2\cos(\omega_2t+\theta_2)-\frac{A_1A_3\epsilon}{2}\{\frac{\cos((\omega_1+\omega_3)t+\theta_1+\theta_3)}{\omega^2_2-(\omega_1+\omega_3)^2}
+\frac{\cos((\omega_1-\omega_3)t+\theta_1-\theta_3)}{\omega^2_2-(\omega_1-\omega_3)^2}\}+O(\epsilon^2),
\end{equation*}
\begin{equation*}
z=A_3\cos(\omega_3t+\theta_3)-\frac{A_1A_2\epsilon}{2}\{\frac{\cos((\omega_1+\omega_2)t+\theta_1+\theta_2)}{\omega^2_3-(\omega_1+\omega_2)^2}
+\frac{\cos((\omega_1-\omega_2)t+\theta_1-\theta_2)}{\omega^2_3-(\omega_1-\omega_2)^2}\}+O(\epsilon^2).
\end{equation*}

\emph{\textbf{Case (ii)}}. $\omega_2+\omega_3=\omega_1$.  Then we
have
\begin{equation*}
x_1=-\frac{A_2A_3}{2}\{(t-t_0)\frac{\sin(\omega_1t+\theta_2+\theta_3)}{2\omega_1}
+\frac{\cos((\omega_2-\omega_3)t+\theta_2-\theta_3)}{\omega^2_1-(\omega_2-\omega_3)^2}\},
\end{equation*}
\begin{equation*}
y_1=-\frac{A_1A_3}{2}\{\frac{\cos((\omega_1+\omega_3)t+\theta_1+\theta_3)}{\omega^2_2-(\omega_1+\omega_3)^2}
+(t-t_0)\frac{\sin(\omega_2t+\theta_1-\theta_3)}{2\omega_2}\},
\end{equation*}
\begin{equation*}
z_1=-\frac{A_1A_2}{2}\{\frac{\cos((\omega_1+\omega_2)t+\theta_1+\theta_2)}{\omega^2_3-(\omega_1+\omega_2)^2}
+(t-t_0)\frac{\sin(\omega_3t+\theta_1-\theta_2)}{2\omega_3}\}.
\end{equation*}
Since there exist secular terms in solutions, we use the TR method
to eliminate them. By expanding $x_0,y_0$ and $z_0$ as the power
series at $t_0$, we take the approximate renormalization equations
as
\begin{equation*}
\frac{\mathrm{d}}{\mathrm{d}t_0}\{A_1(t_0)\cos(\omega_1t_0+\theta_1(t_0))\}=
-\frac{A_2A_3}{4\omega_1}\sin(\omega_1t+\theta_2+\theta_3)-A_1\omega_1\sin(\omega_1t_0+\theta_1),
\end{equation*}
\begin{equation*}
\frac{\mathrm{d}}{\mathrm{d}t_0}\{A_2(t_0)\cos(\omega_2t_0+\theta_2(t_0))\}=
-\frac{A_1A_3}{4\omega_2}\sin(\omega_2t+\theta_1+\theta_3)-A_2\omega_2\sin(\omega_2t_0+\theta_2),
\end{equation*}
\begin{equation*}
\frac{\mathrm{d}}{\mathrm{d}t_0}\{A_3(t_0)\cos(\omega_1t_0+\theta_3(t_0))\}=
-\frac{A_1A_2}{4\omega_3}\sin(\omega_3t+\theta_1+\theta_2)-A_3\omega_3\sin(\omega_3t_0+\theta_3).
\end{equation*}
Correspondingly,  a closed renormalization equations system is taken
as
\begin{equation*}
A_1'=0,\theta_1'=\frac{A_2A_3}{4A_1\omega_1}\epsilon,
\end{equation*}
\begin{equation*}
A_2'=0,\theta_2'=\frac{A_1A_3}{4A_2\omega_2}\epsilon,
\end{equation*}
\begin{equation*}
A_3'=0,\theta_3'=\frac{A_1A_2}{4A_3\omega_3}\epsilon,
\end{equation*}
and a condition
\begin{equation*}
\theta_1=\theta_2+\theta_3,
\end{equation*}
must be satisfied. Under this condition, we give the solutions of
the renormalization equations system as follows
\begin{equation*}
A_1(t_0)=A_0,\theta_1(t_0)=\frac{B_0C_0}{4A_0\omega_1}\epsilon
t_0+\theta_{10},
\end{equation*}
\begin{equation*}
A_2(t_0)=B_0,\theta_2(t_0)=\frac{A_0C_0}{4B_0\omega_2}\epsilon
t_0+\theta_{20},
\end{equation*}
\begin{equation*}
A_3(t_0)=C_0,\theta_3(t_0)=\frac{A_0B_0}{4C_0\omega_3}\epsilon
t_0+\theta_{30},
\end{equation*}
where $A_0,B_0, C_0$ and $\theta_{i0}$ are arbitrary constants for
$i=1,2,3$. Therefore, the asymptotic solutions are
\begin{equation*}
x=A_0\cos((\omega_1+\frac{B_0C_0}{4A_0\omega_1}\epsilon)
t+\theta_{10})
+\epsilon\frac{\cos((\omega_2-\omega_3+\frac{A_0C_0}{4B_0\omega_2}\epsilon
-\frac{A_0B_0}{4C_0\omega_3}\epsilon)
t+\theta_{20}-\theta_{30})}{\omega^2_1-(\omega_2-\omega_3)^2}+O(\epsilon^2),
\end{equation*}
\begin{equation*}
y=B_0\cos((\omega_2+\frac{A_0C_0}{4B_0\omega_2}\epsilon)
t+\theta_{20})
+\epsilon\frac{\cos((\omega_1+\omega_3+\frac{B_0C_0}{4A_0\omega_1}\epsilon
+\frac{A_0B_0}{4C_0\omega_3}\epsilon)
t+\theta_{10}+\theta_{30})}{\omega^2_2-(\omega_1+\omega_3)^2}+O(\epsilon^2),
\end{equation*}
\begin{equation*}
z=C_0\cos((\omega_3+\frac{A_0B_0}{4C_0\omega_3}\epsilon)
t+\theta_{30})
+\epsilon\frac{\cos((\omega_1+\omega_2+\frac{B_0C_0}{4A_0\omega_1}\epsilon
+\frac{A_0C_0}{4B_0\omega_2}\epsilon)
t+\theta_{10}+\theta_{20})}{\omega^2_3-(\omega_1+\omega_2)^2}+O(\epsilon^2).
\end{equation*}

\emph{\textbf{Case (iii)}}. $\omega_2-\omega_3=\omega_1$. By the
similar way in the case (ii), we have the the following asymptotic
solutions
\begin{equation*}
x=A_0\cos((\omega_1+\frac{B_0C_0}{4A_0\omega_1}\epsilon)
t+\theta_{10})
+\epsilon\frac{\cos((\omega_2+\omega_3+\frac{A_0C_0}{4B_0\omega_2}\epsilon
+\frac{A_0B_0}{4C_0\omega_3}\epsilon)
t+\theta_{20}+\theta_{30})}{\omega^2_1-(\omega_2+\omega_3)^2}+O(\epsilon^2),
\end{equation*}
\begin{equation*}
y=B_0\cos((\omega_2+\frac{A_0C_0}{4B_0\omega_2}\epsilon)
t+\theta_{20})
+\epsilon\frac{\cos((\omega_1-\omega_3+\frac{B_0C_0}{4A_0\omega_1}\epsilon
-\frac{A_0B_0}{4C_0\omega_3}\epsilon)
t+\theta_{10}-\theta_{30})}{\omega^2_2-(\omega_1-\omega_3)^2}+O(\epsilon^2),
\end{equation*}
\begin{equation*}
z=C_0\cos((\omega_1+\frac{A_0B_0}{4C_0\omega_3}\epsilon)
t+\theta_{30})
+\epsilon\frac{\cos((\omega_1+\omega_2+\frac{B_0C_0}{4A_0\omega_1}\epsilon
+\frac{A_0C_0}{4B_0\omega_2}\epsilon)
t+\theta_{10}+\theta_{20})}{\omega^2_3-(\omega_1+\omega_2)^2}+O(\epsilon^2).
\end{equation*}

\subsection{A boundary layer problem}
 As a nontrivial  example, we
solve the following singular perturbation equation which had been
discussed by Lighthill(see, for example, [66]) and Tsien[67],
\begin{equation}
(x+\epsilon y)\frac{dy}{dx}+(2+x)y=0, y(1)=A\mathrm{e}^{-1}.
\end{equation}
By the PLK method[66,67], the uniform valid solution at original is
given by
\begin{equation}
y=Ae^{-s}s^{-2}\{1+A\epsilon[\frac{2}{s^3}+\frac{1}{s^3}-\int_s^1e^{-r}
(\frac{2}{r^4}+\frac{1}{r^3})\mathrm{d}r]\}+O(\frac{\epsilon^2}{s^6}),
\end{equation}
where
\begin{equation}
x=s-\frac{\epsilon
A}{3s^2}-\frac{3\epsilon^2A^2}{10s^4}+O(\frac{\epsilon^2}{s^6}).
\end{equation}
This is an implicit form solution represented in terms of
intermediate variable $s$. Now we use the TR method to give an
explicit uniform asymptotic solution. Taking a transformation
$x=\frac{1}{t}$ gives
\begin{equation}
(1+\epsilon ty)\frac{dy}{dt}=\frac{2t+1}{t^2}y,
y(1)=A\mathrm{e}^{-1}.
\end{equation}
Then expanding $y$ as $y=y_0+y_1\epsilon+y_2\epsilon^2+\cdots$ and
substituting it into the above equation yield
\begin{equation}
y_0'=\frac{2t+1}{t^2}y_0,
\end{equation}
\begin{equation}
y_1'=\frac{2t+1}{t^2}y_1-ty_0y_0',
\end{equation}
and so on. Solving these two first equations, we have
\begin{equation}
y_0=A_0t^2e^{-\frac{1}{t}},
\end{equation}
\begin{equation}
y_1=-A^2_0t^2e^{-\frac{1}{t}}\int_{t_0}^t
r(2r+1)e^{-\frac{1}{r}}\mathrm{d}r.
\end{equation}
Therefore, the approximate solution is
\begin{equation}
y=A_0t^2e^{-\frac{1}{t}}-\epsilon A^2_0
t^2e^{-\frac{1}{t}}\int_{t_0}^tr(2r+1)e^{-\frac{1}{r}}\mathrm{d}r,
\end{equation}
which contains a secular term. By the TR method, considering $A_0$
as a function of $t_0$, and the renormalization equation gives
\begin{equation}
A_0'=-\epsilon A^2_0t_0(2t_0+1)e^{-\frac{1}{t_0}},
\end{equation}
whose solution is
\begin{equation}
A_0=\frac{1}{\epsilon\int_{0}^{t_0}r(2r+1)e^{-\frac{1}{r}}\mathrm{d}r+c_0},
\end{equation}
where $c_0$ is a constant. Therefore, the asymptotic solution is
\begin{equation}
y=\frac{t^2e^{-\frac{1}{t}}}{\epsilon\int_{0}^tr(2r+1)e^{-\frac{1}{r}}\mathrm{d}r+c_0}+O(\epsilon^2).
\end{equation}
 By the condition $y(1)=Ae^{-1}$, we get
\begin{equation}
c_0=\frac{1}{A}-\epsilon\int_{0}^1r(2r+1)e^{-\frac{1}{r}}\mathrm{d}r.
\end{equation}
Therefore, the asymptotic solution is
\begin{equation}
y=\frac{t^2e^{-\frac{1}{t}}}{\epsilon\int_{1}^tr(2r+1)e^{-\frac{1}{r}}\mathrm{d}r+\frac{1}{A}}+O(\epsilon^2).
\end{equation}

Next we consider the limitation of $y$ at $t=+\infty$. From the
L'Hopital's rule, we have
\begin{equation}
\lim_{t\rightarrow +\infty}y=\lim_{t\rightarrow
+\infty}\frac{t^2e^{-\frac{1}{t}}}{\epsilon\int_{1}^t
r(2r+1)e^{-\frac{1}{r}}\mathrm{d}r+\frac{1}{A}}=0,
\end{equation}
which means $y$ is a uniform valid asymptotic solution at
$t=+\infty$. In other words,  the uniform asymptotic solution in
terms of $x$
\begin{equation}
y=\frac{x^{-2}e^{-x}}{\epsilon\int_{1}^{\frac{1}{x}}r(2r+1)e^{-\frac{1}{r}}\mathrm{d}r+\frac{1}{A}}+O(\epsilon^2).
\end{equation}
is uniform valid at $x=0$. This solution is explicit and  not
dependent on the intermediate variable which appears in (53) and
(54).

\subsection{Another boundary layer problem}

A similar but more difficult singular perturbation problem is the
following equation
\begin{equation}
(x^2+\epsilon y)\frac{dy}{dx}+y=2x^3+x^2, y(1)=Ae.
\end{equation}
Tsien  pointed out that it can not be solved by PLK method[67].
However, by the TR method, we can give its uniform valid asymptotic
solution at the original. In fact, taking $x=\frac{1}{t}$ and
according to the TR method, we get
\begin{equation}
y=e^t\{A_0+\int_{t_0}^t\frac{1+r}{r^3}e^{-r}dr-\epsilon\int_{t_0}^tr^2e^{-r}y_0(r)y'_0(r)dr\}+O(\epsilon^2),
\end{equation}
where
\begin{equation}
y_0=e^t(A_0+\int_{t_0}^t\frac{1+r}{r^3}e^{-r}dr).
\end{equation}
Supposing that $A_0$ depends on $t_0$, and then an approximate
closed renormalization equation can be taken as
\begin{equation}
A_0'(t_0)=-\epsilon A_0^2(t_0)t^2_0e^{t_0},
\end{equation}
whose solution is
$A_0(t_0)=\frac{1}{\epsilon\int_0^{t_0}r^2e^rdr+c_0}$. Then we write
the final solution as
\begin{equation}
y(x)=\frac{e^{\frac{1}{x}}}{\epsilon\int_1^{\frac{1}{x}}r^2e^rdr+\frac{1}{A}}+O(\epsilon^2),
\end{equation}
which is a uniform valid asymptotic solution in $(0,1)$ since
$\lim_{x\rightarrow 0} y=0$.

\subsection{Boundary value problem in infinite dimension: Asymptotic solution of longitudinal vibration of the rod}

For free longitudinal vibrations of a clamped-clamped rod in
nonlinear elastic medium, the governing equation can be given
by[68-70]
\begin{equation}
ESu_{xx}=\rho Su_{tt}+F(u),
\end{equation}
where $u$ is the longitudinal displacement, $\rho$ is the density of
the rod, $S$ is the area of cross-section, $E$ is Young's modulus,
and $F(u)$ is the restoring force per unit length acting the rod
from the surrounding medium. Furthermore, suppose that the length of
the rod is $l$, and the force $F$ satisfies $F(0)=0$ and
$F(u)=a_1u+a_3u^3+\cdots$. By removing the high order terms of the
expansion of $F(u)$ and re-scaling the variables, we can obtain the
nonlinear wave equation with the boundary and initial conditions
\begin{equation}
u_{tt}-u_{xx}+\mu u=\epsilon u^3,
\end{equation}
\begin{equation}
u(0,t)=u(\pi,t)=0,
\end{equation}
\begin{equation}
u(x,0)=\phi(x), u_t(x,0)=\psi(x),
\end{equation}
where mass parameter $\mu$ is a positive real number, and $\epsilon$
is a positive small parameter representing a perturbation term.

 This
equation also arises from the nonlinear vibration of continuous
structures such as beams, plates and shells, etc. Since there exist
infinite degree of freedom, the phenomena of internal resonances
will be complex, and it will causes the coupling of normal modes.
This is a perturbation equation with time variable, a natural
problem is to study its asymptotic behavior at large time. A lot of
perturbation technics an be used to deal with this kind of the
problems[66]. Because the continuous structure has infinite degree
of freedom, by the usages of the routine perturbation methods, we
have to take a truncation procedure to deal with the
infinite-dimensional algebraic or differential equations system. In
many cases, only few modes can be represented which will cause
significants errors[71]. Andrianov and Danishevskyy used an
artificial small parameter method and numerical method to construct
the asymptotic solution to the vibrations of nonlinear structures
and give the analysis of the resonance interactions between
modes[72]. Here we use the TR method to give it an analytic
asymptotic solution.

Now we  consider asymptotic solution of the nonlinear wave equation
(74). Expanding $u$ as a power series of the small parameter
$\epsilon$
\begin{equation}
u=u_0+u_1\epsilon+u_2\epsilon^2+\cdots,
\end{equation}
where every $u_n$ satisfies the boundary condition, and substituting
it into the above equation gives
\begin{equation}
u_{0tt}-u_{0xx}+\mu u_0=0,
\end{equation}
\begin{equation}
u_{1tt}-u_{1xx}+\mu u_1=u_0^3.
\end{equation}
By the boundary condition, we can expand the solution $u_0$ as
\begin{equation}
u_0=\sum_{n=1}^{+\infty}A_n\cos(\omega_nt+\theta_n)\sin(nx),
\end{equation}
where $\omega_n=\sqrt{n^2+\mu}$. Correspondingly, the equation of
$u_1$ becomes
\begin{equation}
u_{1tt}-u_{1xx}+\mu
u_1=(\sum_{m=1}^{+\infty}A_m\cos(\omega_mt+\theta_m)\sin(mx))^3.
\end{equation}
By the boundary condition, we can expand $u_1$ as
\begin{equation}
u_1=\sum_{n=1}^{+\infty}v_n(t)\sin(nx).
\end{equation}
At the same time, we can expand the right side of the Eq.(81) as
\begin{equation}
(\sum_{m=1}^{+\infty}A_m\cos(\omega_mt+\theta_m)\sin(mx))^3=\sum_{n=1}^{+\infty}H_n(t)\sin(nx).
\end{equation}
Therefore, we get the infinite-dimensional equations system
\begin{equation}
v''_n(t)+\omega^2_nv_n(t)=H_n(t), n=1,2,\cdots.
\end{equation}

In order to get $H_n(t)$, we need some complicated computations
including expansion of three multiple series. In fact, we have
\begin{equation*}
H_1(t)=\frac{3}{4}\sum_{k=1}^{+\infty}\sum_{r=1}^{k}A_{1+k-r}A_kA_r\cos(\omega_{1+k-r}t+\theta_{1+k-r})
\cos(\omega_{k}t+\theta_{k})\cos(\omega_{r}t+\theta_{r})
\end{equation*}
\begin{equation}
-\frac{3}{4}\sum_{k=3}^{+\infty}\sum_{r=1}^{k-2}A_{k-r-1}A_kA_r\cos(\omega_{k-r-1}t+\theta_{k-r-1})
\cos(\omega_{k}t+\theta_{k})\cos(\omega_{r}t+\theta_{r});
\end{equation}
\begin{equation*}
H_2(t)=\frac{3}{4}\sum_{k=1}^{+\infty}\sum_{r=1}^{k+1}A_{1+k-r}A_kA_r\cos(\omega_{2+k-r}t+\theta_{2+k-r})
\cos(\omega_{k}t+\theta_{k})\cos(\omega_{r}t+\theta_{r})
\end{equation*}
\begin{equation}
-\frac{3}{4}\sum_{k=4}^{+\infty}\sum_{r=1}^{k-3}A_{k-r-2}A_kA_r\cos(\omega_{k-r-2}t+\theta_{2+k-r})
\cos(\omega_{k}t+\theta_{k})\cos(\omega_{r}t+\theta_{r});
\end{equation}
\begin{equation*}
H_n(t)=\frac{3}{4}\sum_{k=1}^{+\infty}\sum_{r=1}^{n+k-1}A_{n+k-r}A_kA_r\cos(\omega_{n+k-r}t+\theta_{n+k-r})
\cos(\omega_{k}t+\theta_{k})\cos(\omega_{r}t+\theta_{r})
\end{equation*}
\begin{equation*}
-\frac{3}{4}\sum_{k={n+2}}^{+\infty}\sum_{r=1}^{k-n-1}A_{k-n-r}A_kA_r\cos(\omega_{k-n-r}t+\theta_{k-n-r})
\cos(\omega_{k}t+\theta_{k})\cos(\omega_{r}t+\theta_{r})
\end{equation*}
\begin{equation}
-\frac{1}{4}\sum_{k=1}^{n-2}\sum_{r=1}^{n-k-1}A_{n-k-r}A_{k}A_r\cos(\omega_{n-k-r}t+\theta_{n-k-r})
\cos(\omega_{k}t+\theta_{k})\cos(\omega_{r}t+\theta_{r}),
\end{equation}
for $n\geq 3$. Furthermore, we have
\begin{equation*}
\sum_{k=1}^{+\infty}\sum_{r=1}^{n+k-1}A_{n+k-r}A_kA_r\cos(\omega_{n+k-r}t+\theta_{n+k-r})
\cos(\omega_{k}t+\theta_{k})\cos(\omega_{r}t+\theta_{r})
\end{equation*}
\begin{equation*}
=(\sum_{k=1}^{+\infty}A^2_k)A_n\cos(\omega_nt+\theta_n)+\frac{1}{4}A^3_n\cos3(\omega_nt+\theta_n)
\end{equation*}
\begin{equation*}
+\frac{1}{4}\sum_{k=1}^{+\infty}\sum_{r=1,r\neq
k}^{n+k-1}A_{n+k-r}A_kA_r\cos((\omega_{n+k-r}+\omega_k-\omega_r)t+\theta_{n+k-r}+\theta_k-\theta_r)
\end{equation*}
\begin{equation*}
+\frac{1}{4}\sum_{k=1}^{+\infty}\sum_{r=1,r\neq
n}^{n+k-1}A_{n+k-r}A_kA_r\cos((\omega_k+\omega_r-\omega_{n+k-r})t+\theta_k+\theta_r-\theta_{n+k-r})
\end{equation*}
\begin{equation*}
+\frac{1}{4}\sum_{k=1}^{+\infty}\sum_{r=1,r\neq k,r\neq
n}^{n+k-1}A_{n+k-r}A_kA_r\cos((\omega_{n+k-r}+\omega_r-\omega_k)t+\theta_{n+k-r}+\theta_r-\theta_k)
\end{equation*}
\begin{equation*}
+\frac{1}{4}\sum_{k=1,k\neq
n}^{+\infty}\sum_{r=1}^{n+k-1}A_{n+k-r}A_kA_r\cos((\omega_{n+k-r}+\omega_r+\omega_k)t+\theta_{n+k-r}+\theta_r+\theta_k)
\end{equation*}
\begin{equation}
+\frac{1}{4}\sum_{k=1}^{+\infty}\sum_{r=1,r\neq
n}^{n+k-1}A_{n+k-r}A_kA_r\cos((\omega_{n+k-r}+\omega_r+\omega_k)t+\theta_{n+k-r}+\theta_r+\theta_k);
\end{equation}
and
\begin{equation*}
\sum_{k={n+2}}^{+\infty}\sum_{r=1}^{k-n-1}A_{k-n-r}A_kA_r\cos(\omega_{k-n-r}t+\theta_{k-n-r})
\cos(\omega_{k}t+\theta_{k})\cos(\omega_{r}t+\theta_{r})
\end{equation*}
\begin{equation*}
=\frac{1}{4}\sum_{k={n+2}}^{+\infty}\sum_{r=1}^{k-n-1}A_{n-k-r}A_{k}A_r\cos((\omega_{n-k-r}+\omega_{k}
-\omega_r)t+\theta_{n-k-r} +\theta_{k}-\theta_r)
\end{equation*}
\begin{equation*}
+\frac{1}{4}\sum_{k={n+2}}^{+\infty}\sum_{r=1}^{k-n-1}A_{n-k-r}A_{k}A_r\cos((\omega_{n-k-r}+\omega_{r}-\omega_{k})t+\theta_{n-k-r}
+\theta_r-\theta_{k})
\end{equation*}
\begin{equation*}
+\frac{1}{4}\sum_{k={n+2}}^{+\infty}\sum_{r=1}^{k-n-1}A_{n-k-r}A_{k}A_r\cos((\omega_{k}+\omega_r-\omega_{n-k-r})t+
\theta_{k}+\theta_r-\theta_{n-k-r})
\end{equation*}
\begin{equation}
+\frac{1}{4}\sum_{k={n+2}}^{+\infty}\sum_{r=1}^{k-n-1}A_{n-k-r}A_{k}A_r\cos((\omega_{n-k-r}+\omega_{k}+\omega_r)t+\theta_{n-k-r}
+\theta_{k}+\theta_r),
\end{equation}
for $n\geq1$, and
\begin{equation*}
\sum_{k=1}^{n-2}\sum_{r=1}^{n-k-1}A_{n-k-r}A_{k}A_r\cos(\omega_{n-k-r}t+\theta_{n-k-r})
\cos(\omega_{k}t+\theta_{k})\cos(\omega_{r}t+\theta_{r})
\end{equation*}
\begin{equation*}
=\frac{1}{4}\sum_{k=1}^{n-2}\sum_{r=1}^{n-k-1}A_{n-k-r}A_{k}A_r\cos((\omega_{n-k-r}+\omega_{k}
-\omega_r)t+\theta_{n-k-r} +\theta_{k}-\theta_r)
\end{equation*}
\begin{equation*}
+\frac{1}{4}\sum_{k=1}^{n-2}\sum_{r=1}^{n-k-1}A_{n-k-r}A_{k}A_r\cos((\omega_{n-k-r}+\omega_{r}-\omega_{k})t+\theta_{n-k-r}
+\theta_r-\theta_{k})
\end{equation*}
\begin{equation*}
+\frac{1}{4}\sum_{k=1}^{n-2}\sum_{r=1}^{n-k-1}A_{n-k-r}A_{k}A_r\cos((\omega_{k}+\omega_r-\omega_{n-k-r})t+
\theta_{k}+\theta_r-\theta_{n-k-r})
\end{equation*}
\begin{equation}
+\frac{1}{4}\sum_{k=1}^{n-2}\sum_{r=1}^{n-k-1}A_{n-k-r}A_{k}A_r\cos((\omega_{n-k-r}+\omega_{k}+\omega_r)t+\theta_{n-k-r}
+\theta_{k}+\theta_r).
\end{equation}
From the above formulas, we know that there exists only one
resonance term
$(\sum_{k=1}^{+\infty}A^2_k)A_n\cos(\omega_nt+\theta_n)$ for each
mode $v_n$ when $\mu\neq0$, and this resonance term will lead to a
secular term correspondingly. In fact, according to the expressions
of $H_n$, we can easily give the exact solution for every $v_n$ by
the following formulas,
\begin{equation}
v_n=-\frac{\epsilon}{4\omega_n}(\sum_{k=1}^{+\infty}A^2_k)A_n(t-t_0)\sin(\omega_nt+\theta_n)-
\frac{9\epsilon}{128\omega_n^2}A^3_n\cos3(\omega_nt+\theta_n)+\frac{3}{4}I_n\epsilon-\frac{3}{4}J_n\epsilon,
n=1,2;
\end{equation}
\begin{equation*}
v_n=-\frac{\epsilon}{4\omega_n}(\sum_{k=1}^{+\infty}A^2_k)A_n(t-t_0)\sin(\omega_nt+\theta_n)-
\frac{9\epsilon}{128\omega_n^2}A^3_n\cos3(\omega_nt+\theta_n)+\frac{3}{4}I_n\epsilon-\frac{3}{4}J_n\epsilon-\frac{1}{4}K_n\epsilon,
\end{equation*}
\begin{equation}
n\geq3,
\end{equation}
where
\begin{equation*}
I_n=\frac{1}{4}\sum_{k=1}^{+\infty}\sum_{r=1,r\neq
k}^{n+k-1}\frac{A_{n+k-r}A_kA_r}{\omega^2_n-(\omega_{n+k-r}+\omega_k-\omega_r)^2}
\cos((\omega_{n+k-r}+\omega_k-\omega_r)t+\theta_{n+k-r}+\theta_k-\theta_r)
\end{equation*}
\begin{equation*}
+\frac{1}{4}\sum_{k=1}^{+\infty}\sum_{r=1,r\neq
n}^{n+k-1}\frac{A_{n+k-r}A_kA_r}{\omega^2_n-(\omega_k+\omega_r-\omega_{n+k-r})^2}
\cos((\omega_k+\omega_r-\omega_{n+k-r})t+\theta_k+\theta_r-\theta_{n+k-r})
\end{equation*}
\begin{equation*}
+\frac{1}{4}\sum_{k=1}^{+\infty}\sum_{r=1,r\neq k,r\neq
n}^{n+k-1}\frac{A_{n+k-r}A_kA_r}{\omega^2_n-(\omega_{n+k-r}+\omega_r-\omega_k)^2}
\cos((\omega_{n+k-r}+\omega_r-\omega_k)t+\theta_{n+k-r}+\theta_r-\theta_k)
\end{equation*}
\begin{equation*}
+\frac{1}{4}\sum_{k=1,k\neq
n}^{+\infty}\sum_{r=1}^{n+k-1}\frac{A_{n+k-r}A_kA_r}{\omega^2_n-(\omega_{n+k-r}+\omega_r+\omega_k)^2}
\cos((\omega_{n+k-r}+\omega_r+\omega_k)t+\theta_{n+k-r}+\theta_r+\theta_k)
\end{equation*}
\begin{equation}
+\frac{1}{4}\sum_{k=1}^{+\infty}\sum_{r=1,r\neq
n}^{n+k-1}A_{n+k-r}A_kA_r\cos((\omega_{n+k-r}+\omega_r+\omega_k)t+\theta_{n+k-r}+\theta_r+\theta_k),
\end{equation}
\begin{equation*}
J_n=\frac{1}{4}\sum_{k=n+2}^{+\infty}\sum_{r=1}^{k-n-1}\frac{A_{n-k-r}A_{k}A_r}{\omega^2_n-(
\omega_{n-k-r}+\omega_{k}-\omega_r)^2}
\cos((\omega_{n-k-r}+\omega_{k}-\omega_r)t+\theta_{n-k-r}
+\theta_{k}-\theta_r)
\end{equation*}
\begin{equation*}
+\frac{1}{4}\sum_{k=n+2}^{+\infty}\sum_{r=1}^{k-n-1}\frac{A_{n-k-r}A_{k}A_r}{\omega^2_n-
(\omega_{n-k-r}+\omega_{r}-\omega_{k})^2}
\cos((\omega_{n-k-r}+\omega_{r}-\omega_{k})t+\theta_{n-k-r}
+\theta_r-\theta_{k})
\end{equation*}
\begin{equation*}
+\frac{1}{4}\sum_{k=n+2}^{+\infty}\sum_{r=1}^{k-n-1}\frac{A_{n-k-r}A_{k}A_r}{\omega^2_n-
(\omega_{k}+\omega_r-\omega_{n-k-r})^2}
\cos((\omega_{k}+\omega_r-\omega_{n-k-r})t+
\theta_{k}+\theta_r-\theta_{n-k-r})
\end{equation*}
\begin{equation}
+\frac{1}{4}\sum_{k=n+2}^{+\infty}\sum_{r=1}^{k-n-1}\frac{A_{n-k-r}A_{k}A_r}{\omega^2_n-
(\omega_{n-k-r}+\omega_{k-r}+\omega_r)^2}
\cos((\omega_{n-k}+\omega_{k}+\omega_r)t+\theta_{n-k-r}
+\theta_{k}+\theta_r),
\end{equation}
and
\begin{equation*}
K_n=\frac{1}{4}\sum_{k=2}^{n-1}\sum_{r=1}^{k-1}\frac{A_{n-k-r}A_{k}A_r}{\omega^2_n-(
\omega_{n-k-r}+\omega_{k}-\omega_r)^2}
\cos((\omega_{n-k-r}+\omega_{k}-\omega_r)t+\theta_{n-k-r}
+\theta_{k}-\theta_r)
\end{equation*}
\begin{equation*}
+\frac{1}{4}\sum_{k=2}^{n-1}\sum_{r=1}^{k-1}\frac{A_{n-k-r}A_{k}A_r}{\omega^2_n-
(\omega_{n-k-r}+\omega_{r}-\omega_{k})^2}
\cos((\omega_{n-k-r}+\omega_{r}-\omega_{k})t+\theta_{n-k-r}
+\theta_r-\theta_{k})
\end{equation*}
\begin{equation*}
+\frac{1}{4}\sum_{k=2}^{n-1}\sum_{r=1}^{k-1}\frac{A_{n-k-r}A_{k}A_r}{\omega^2_n-
(\omega_{k}+\omega_r-\omega_{n-k-r})^2}
\cos((\omega_{k}+\omega_r-\omega_{n-k-r})t+
\theta_{k}+\theta_r-\theta_{n-k-r})
\end{equation*}
\begin{equation}
+\frac{1}{4}\sum_{k=2}^{n-1}\sum_{r=1}^{k-1}\frac{A_{n-k-r}A_{k}A_r}{\omega^2_n-
(\omega_{n-k-r}+\omega_{k}+\omega_r)^2}
\cos((\omega_{n-k-r}+\omega_{k}+\omega_r)t+\theta_{n-k-r}
+\theta_{k}+\theta_r).
\end{equation}
Furthermore, by writing the solution of $u$ as a Fourier's series
\begin{equation}
u(x,t)=\sum_{n=1}^{+\infty}U_n(t,\epsilon)\sin nx,
\end{equation}
we have
\begin{equation*}
U_n=A_n\cos(\omega_nt+\theta_n)-\frac{\epsilon}{4\omega_n}(\sum_{k=1}^
{+\infty}A^2_k)A_n(t-t_0)\sin(\omega_nt+\theta_n)
\end{equation*}
\begin{equation}
-\frac{9\epsilon}{128\omega_n^2}A^3_n\cos3(\omega_nt+\theta_n)+\frac{3}{4}I_n\epsilon-\frac{3}{4}J_n\epsilon
+O(\epsilon^2), n=1,2;
\end{equation}
\begin{equation*}
U_n=A_n\cos(\omega_nt+\theta_n)-\frac{\epsilon}{4\omega_n}(\sum_{k=1}^
{+\infty}A^2_k)A_n(t-t_0)\sin(\omega_nt+\theta_n)
\end{equation*}
\begin{equation}
-\frac{9\epsilon}{128\omega_n^2}A^3_n\cos3(\omega_nt+\theta_n)+\frac{3}{4}I_n\epsilon-\frac{3}{4}J_n\epsilon
-\frac{1}{4}K_n\epsilon+O(\epsilon^2), n\geq3.
\end{equation}
By considering $A_n$ and $\theta_n$ as the functions of $t_0$ and
neglecting the terms of $I_n$, $J_n$ and $K_n$, the renormalization
equations are taken as
\begin{equation*}
(A_n\cos(\omega_nt_0+\theta_n)-\frac{9\epsilon}{128\omega_n^2}A^3_n\cos3(\omega_nt_0+\theta_n))'
=-\frac{\epsilon}{4\omega_n}(\sum_{k=1}^{+\infty}A^2_k)A_n\sin(\omega_nt_0+\theta_n)
\end{equation*}
\begin{equation}
-A_n\omega_n\sin(\omega_nt_0+\theta_n)+\frac{27\epsilon}{128\omega_n}A^3_n\sin3(\omega_nt_0+\theta_n),
n=1,2,\cdots,
\end{equation}
where the prime is the derivative with respect to $t_0$. In order to
find the solutions of $A_n$ and $\theta_n$, we take a closed
approximate equations system
\begin{equation}
A'_n=0,\theta'_n=\frac{\epsilon}{4\omega_n}(\sum_{k=1}^{+\infty}A^2_k),
n=1,2,\cdots,
\end{equation}
whose solutions are
\begin{equation}
A_n(t_0)=A_{n0},\theta_n(t_0)=\frac{\epsilon}{4\omega_n}(\sum_{k=1}^{+\infty}A^2_{k0})t_0+\theta_{n0},
n=1,2,\cdots,
\end{equation}
where $A_{n0}$ and $\theta_{n0}$ are constants which can be
determined by initial conditions. This choice of the closed
renormalization equations system is reasonable since the amplitudes
are slow variables and hence can be taken as constants. Then the
asymptotic solutions of $U_n's$ are given by
\begin{equation*}
U_n(t)=A_{n0}\cos((\omega_n+\frac{\epsilon}{4\omega_n}(\sum_{k=1}^{+\infty}A^2_{k0}))t+\theta_{n0})
\end{equation*}
\begin{equation}
-\frac{9\epsilon}{128\omega_n^2}A^3_{n0}\cos3((\omega_n+\frac{\epsilon}{4\omega_n}
(\sum_{k=1}^{+\infty}A^2_{k0}))t+\theta_{n0})+E_n,
\end{equation}
where $E_n$ represents the terms including $I_n$, $J_n$ and $K_n$ in
which $A_n$ and $\theta_n$ must be replaced by their solutions
(101), and then the asymptotic solution of $u$ is
\begin{equation*}
u(x,t)=\sum_{n=1}^{+\infty}\{A_{n0}\cos((\omega_n+\frac{\epsilon}{4\omega_n}(\sum_{k=1}^{+\infty}A^2_{k0}))t
+\theta_{n0})
\end{equation*}
\begin{equation}
-\frac{9\epsilon}{128\omega_n^2}A^3_{n0}\cos3((\omega_n+\frac{\epsilon}{4\omega_n}
(\sum_{k=1}^{+\infty}A^2_{k0}))t+\theta_{n0})+E_n\}\sin nx.
\end{equation}
From the above solution, if we only take the first two terms in
$U_n$, we know that the period of $n$-th mode $U_n$ is
\begin{equation}
T_n=\frac{2\pi}{\omega_n+\frac{\epsilon}{4\omega_n}(\sum_{k=1}^{+\infty}A^2_{k0})},n=1,2,\cdots.
\end{equation}
It is also easy to see that these periods  are in general
Diophantine independence, so the asymptotic solution is
quasi-periodic solution.

Next we discuss the case of $\mu=0$. In this case, we have
$\omega_k=k$, so the fourth term in the right side of Eq.(88) and
the last terms in the right side of Eqs.(89-90) become three
resonance terms and hence lead to the secular terms. We write these
formulas as follows

\begin{equation*}
v_n=-\frac{\epsilon}{4\omega_n}(\sum_{k=1}^{+\infty}A^2_k)A_n(t-t_0)\sin(\omega_nt+\theta_n)-
\frac{9\epsilon}{128\omega_n^2}A^3_n\cos3(\omega_nt+\theta_n)
\end{equation*}
\begin{equation*}
+\frac{3\epsilon}{32\omega_n}\sum_{k=1}^{+\infty}\sum_{r=1,r\neq
k,r\neq
n}^{n+k-1}A_{n+k-r}A_kA_r(t-t_0)\sin(\omega_nt+\theta_{n+k-r}+\theta_r-\theta_k)
+\frac{3}{4}I^*_n\epsilon-\frac{3}{4}J^*\epsilon
\end{equation*}
\begin{equation*}
-\frac{3\epsilon}{32\omega_n}\sum_{k=n+2}^{+\infty}\sum_{r=1}^{n-k-1}A_{n+k-r}A_kA_r(t-t_0)\sin(\omega_nt+\theta_{n+k-r}+\theta_r-\theta_k)
+\frac{3}{4}I^*_n\epsilon-\frac{3}{4}J^*\epsilon
\end{equation*}
\begin{equation}
n=1,2;
\end{equation}
\begin{equation*}
v_n=-\frac{\epsilon}{4\omega_n}(\sum_{k=1}^{+\infty}A^2_k)A_n(t-t_0)\sin(\omega_nt+\theta_n)-
\frac{9\epsilon}{128\omega_n^2}A^3_n\cos3(\omega_nt+\theta_n)
\end{equation*}
\begin{equation*}
+\frac{3\epsilon}{32\omega_n}\sum_{k=1}^{+\infty}\sum_{r=1,r\neq
k,r\neq
n}^{n+k-1}A_{n+k-r}A_kA_r(t-t_0)\sin(\omega_nt+\theta_{n+k-r}+\theta_r-\theta_k)
\end{equation*}
\begin{equation*}
-\frac{3\epsilon}{32\omega_n}\sum_{k=n+2}^{+\infty}\sum_{r=1}^{n-k-1}A_{n+k-r}A_kA_r(t-t_0)
\sin(\omega_nt+\theta_{n+k-r}+\theta_r-\theta_k)
\end{equation*}
\begin{equation*}
-\frac{\epsilon}{32\omega_n}\sum_{k=2}^{n-1}\sum_{r=1}^{k-1}A_{n+k}A_{k-r}A_r(t-t_0)
\sin(\omega_nt+\theta_{n-k}+\theta_{k-r}+\theta_r)
+\frac{3}{4}I^*_n\epsilon-\frac{3}{4}J^*_n\epsilon-\frac{1}{4}K^*_n\epsilon,
\end{equation*}
\begin{equation}
n\geq3,
\end{equation}
where
\begin{equation*}
I^*_n=\frac{1}{4}\sum_{k=1}^{+\infty}\sum_{r=1,r\neq
k}^{n+k-1}\frac{A_{n+k-r}A_kA_r}{\omega^2_n-(\omega_{n+k-r}+\omega_k-\omega_r)^2}
\cos((\omega_{n+k-r}+\omega_k-\omega_r)t+\theta_{n+k-r}+\theta_k-\theta_r)
\end{equation*}
\begin{equation*}
+\frac{1}{4}\sum_{k=1}^{+\infty}\sum_{r=1,r\neq
n}^{n+k-1}\frac{A_{n+k-r}A_kA_r}{\omega^2_n-(\omega_k+\omega_r-\omega_{n+k-r})^2}
\cos((\omega_k+\omega_r-\omega_{n+k-r})t+\theta_k+\theta_r-\theta_{n+k-r})
\end{equation*}
\begin{equation*}
+\frac{1}{4}\sum_{k=1,k\neq
n}^{+\infty}\sum_{r=1}^{n+k-1}\frac{A_{n+k-r}A_kA_r}{\omega^2_n-(\omega_{n+k-r}+\omega_r+\omega_k)^2}
\cos((\omega_{n+k-r}+\omega_r+\omega_k)t+\theta_{n+k-r}+\theta_r+\theta_k)
\end{equation*}
\begin{equation}
+\frac{1}{4}\sum_{k=1}^{+\infty}\sum_{r=1,r\neq
n}^{n+k-1}A_{n+k-r}A_kA_r\cos((\omega_{n+k-r}+\omega_r+\omega_k)t+\theta_{n+k-r}+\theta_r+\theta_k);
\end{equation}
\begin{equation*}
J^*_n=\frac{1}{4}\sum_{k=n+2}^{+\infty}\sum_{r=1}^{n-k-1}\frac{A_{n-k-r}A_{k}A_r}{\omega^2_n-(
\omega_{n-k-r}+\omega_{k}-\omega_r)^2}
\cos((\omega_{n-k-r}+\omega_{k}-\omega_r)t+\theta_{n-k-r}
+\theta_{k}-\theta_r)
\end{equation*}
\begin{equation*}
+\frac{1}{4}\sum_{k=n+2}^{+\infty}\sum_{r=1}^{n-k-1}\frac{A_{n-k-r}A_{k}A_r}{\omega^2_n-
(\omega_{n-k-r}+\omega_{r}-\omega_{k})^2}
\cos((\omega_{n-k-r}+\omega_{r}-\omega_{k})t+\theta_{n-k-r}
+\theta_r-\theta_{k})
\end{equation*}
\begin{equation}
+\frac{1}{4}\sum_{k=n+2}^{+\infty}\sum_{r=1}^{n-k-1}\frac{A_{n-k-r}A_{k}A_r}{\omega^2_n-
(\omega_{k}+\omega_r-\omega_{n-k-r})^2}
\cos((\omega_{k}+\omega_r-\omega_{n-k-r})t+
\theta_{k}+\theta_r-\theta_{n-k-r}),
\end{equation}
and
\begin{equation*}
K^*_n=\frac{1}{4}\sum_{k=1}^{n-2}\sum_{r=1}^{k-1}\frac{A_{n-k-r}A_{k}A_r}{\omega^2_n-(
\omega_{n-k-r}+\omega_{k}-\omega_r)^2}
\cos((\omega_{n-k-r}+\omega_{k}-\omega_r)t+\theta_{n-k-r}
+\theta_{k}-\theta_r)
\end{equation*}
\begin{equation*}
+\frac{1}{4}\sum_{k=1}^{n-2}\sum_{r=1}^{k-1}\frac{A_{n-k-r}A_{k}A_r}{\omega^2_n-
(\omega_{n-k-r}+\omega_{r}-\omega_{k})^2}
\cos((\omega_{n-k-r}+\omega_{r}-\omega_{k})t+\theta_{n-k-r}
+\theta_r-\theta_{k})
\end{equation*}
\begin{equation}
+\frac{1}{4}\sum_{k=1}^{n-2}\sum_{r=1}^{k-1}\frac{A_{n-k-r}A_{k}A_r}{\omega^2_n-
(\omega_{k}+\omega_r-\omega_{n-k-r})^2}
\cos((\omega_{k}+\omega_r-\omega_{n-k-r})t+
\theta_{k}+\theta_r-\theta_{n-k-r}).
\end{equation}

By the TR method, we can eliminate these secular terms and give the
corresponding asymptotic solution as follows,

\begin{equation*}
u(x,t)=\sum_{n=1}^{+\infty}\{A_{n0}\cos((n+\frac{\epsilon}{4n}(\sum_{k=1}^{+\infty}A^2_{k0}))t
+\theta_{n0})
\end{equation*}
\begin{equation}
-\frac{9\epsilon}{128n^2}A^3_{n0}\cos3((n+\frac{\epsilon}{4n}
(\sum_{k=1}^{+\infty}A^2_{k0}))t+\theta_{n0})+E_n^*\}\sin nx,
\end{equation}
where $E^*_n$ represents the terms including $I^*_n$, $J^*_n$ and
$K^*_n$ in which $A_n$ and $\theta_n$ must also be replaced by their
solutions (101).

As the most simply but enough nice approximation, we can take the
first term as an asymptotic solution
\begin{equation*}
u(x,t)\simeq
\sum_{n=1}^{+\infty}\{A_{n0}\cos((n+\frac{\epsilon}{4n}(\sum_{k=1}^{+\infty}A^2_{k0}))t
+\theta_{n0})
\end{equation*}
\begin{equation}
-\frac{9\epsilon}{128n^2}A^3_{n0}\cos3((n+\frac{\epsilon}{4n}
(\sum_{k=1}^{+\infty}A^2_{k0}))t+\theta_{n0})\}\sin nx,
\end{equation}
in which each mode is a periodic function and the period of the
$n$-th mode is $2\pi/(n+\frac{\epsilon}{4n}
(\sum_{k=1}^{+\infty}A^2_{k0}))$ which is the $\epsilon$-order
correction of the case of free vibration.

\textbf{Remark 2}. Although the computation for the problem seems to
be very complicated, the renormalization part is very simple. This
is firstly because the problem is with the boundary conditions so
that it becomes an infinite-dimensional ordinary differential
equations system, and secondly, the cubic nonlinear term causes the
complicated  trigonometric functions expansions. Comparing our
result and method with the results obtained by artificial small
parameter method in [72], we can see that the TR method is more
simple and direct, and the result obtained is an analytic asymptotic
solution.

\subsection{Lateral vibrations of the beam}
 We consider the asymptotic solutions of
lateral vibrations of the beam. If the influence of the dynamical
axial force $\frac{ES}{2l}w_{xx}\int_0^lw^2_x\mathrm{d}x$ is
neglected, the governing equation of free vibration of a simple
supported beam on a nonlinear elastic foundation can be given as
follows,
\begin{equation}
ELw_{xxxx}+\rho Sw_{tt}+F(w)=0,
\end{equation}
where $w$ is the lateral displacement, $E$ is the Young's modulus,
$\rho$ is the density of the beam, $S$ is the area of the cross
section,  and $I$ is the moment of inertia of the cross section.
Furthermore, the nonlinear restoring force per unit length is
assumed to be a odd function and then has the following form
$F(w)=aw+bw^3+\cdots.$ By re-scaling of $t,x$ and $w$, we can get a
forth order equation
\begin{equation}
u_{tt}+u_{xxxx}-\alpha u=\epsilon u^3,
\end{equation}
where $\alpha>0$ and $\epsilon\ll 1$ is a small parameter. The
corresponding initial and boundary conditions are
\begin{equation}
u(0,t)=u(2\pi,t)=0,
\end{equation}
\begin{equation}
u(x,0)=\psi(t), u_t(0,x)=\phi(t).
\end{equation}
This is a regular perturbation problem, and it has been discussed by
the artificial small parameter approach in [72]. Here we use the
same TR method to solve it. Expanding $u$ as a power series of the
small parameter $\epsilon$
\begin{equation}
u=u_0+u_1\epsilon+u_2\epsilon^2+\cdots,
\end{equation}
where every $u_n$ satisfies the boundary condition, and substituting
it into the above equation (113) gives
\begin{equation}
u_{0tt}+u_{0xxxx}-\alpha u_0=0,
\end{equation}
\begin{equation}
u_{1tt}+u_{1xxxxx}-\alpha u_1=u_0^3.
\end{equation}
By the boundary condition, we can expand the solution $u_0$ as
\begin{equation}
u_0=\sum_{n=1}^{+\infty}A_n\cos(\omega_nt+\theta_n)\sin(nx),
\end{equation}
where $\omega_n=\sqrt{n^4-\alpha}$. By the same method in the above
example, the asymptotic solution of $u$ is given by
\begin{equation*}
u(x,t)=\sum_{n=1}^{+\infty}\{A_{n0}\cos((\omega_n+\frac{\epsilon}{4\omega_n}(\sum_{k=1}^{+\infty}A^2_{k0}))t
+\theta_{n0})
\end{equation*}
\begin{equation}
-\frac{9\epsilon}{128\omega_n^2}A^3_{n0}\cos3((\omega_n+\frac{\epsilon}{4\omega_n}
(\sum_{k=1}^{+\infty}A^2_{k0}))t+\theta_{n0})+E_n\}\sin nx.
\end{equation}
This is a global valid analytic asymptotic solution.

\subsection{On normal form theory and reduction of for perturbation differential
equation and dynamical systems}

In [42], DeVille et al use the RG method to study the normal form of
the perturbation differential equations. For the initial value
problem of the equation
\begin{equation}
\frac{dx}{dt}=Ax+\epsilon f(x),
\end{equation}
\begin{equation}
x(t_0)=w(t_0),
\end{equation}
where $f(x)=\sum_{\alpha,i}C_{\alpha,i}x^{\alpha}e_i$, $\alpha$ is a
multi-index, $i$ runs from 1 to $n$, $e_i$ is the standard Euclidean
basis vector, the summation is finite, $t_0$ is initial time, and
$\epsilon$ is a positive parameter. Furthermore, Assume that the
matrix $A$ is diagonal. In [42], by taking
\begin{equation}
w(t_0)=W(t_0)+\sum_{\Lambda_{\alpha,i}\neq
0}\frac{C_{\alpha,i}}{\Lambda_{\alpha,i}}W^{\alpha}(t_0)e_i,
\end{equation}
the RG equation in RG method is just
\begin{equation}
\frac{dW}{dt}=AW+\sum_{\Lambda_{\alpha,i}=
0}\frac{C_{\alpha,i}}{\Lambda_{\alpha,i}}W^{\alpha}e_i.
\end{equation}
This is precisely the normal form of Eq.(121) up to and including
$O(\epsilon)$. In [42], the authors used a long discussion and
computation to get the conclusion. However, based on our theory,
this is a simple thing. Firstly, the solution of the Eq.(121) is
\begin{equation*}
x(t)=e^{A(t-t_0)}w(t_0)+\epsilon(t-t_0)e^{A(t-t_0)}\sum_{\Lambda_{\alpha,i}=
0}C_{\alpha,i}w^{\alpha}(t_0)e_i
\end{equation*}
\begin{equation}
+\epsilon e^{A(t-t_0)}\sum_{\Lambda_{\alpha,i}\neq
0}\frac{C_{\alpha,i}}{\Lambda_{\alpha,i}}w^{\alpha}(t_0)e_i+O(\epsilon
^2).
\end{equation}
Now we use the TR method to direct get the result of DeVille et al.
In fact, since $w(t_0)$ is just the solution of $x$ if $t_0$ is
considered as the parameter, $w$ satisfies the same equation with
$x$. So we only need to substituting the variable transformation
(123) into (121) in which $x$ is replaced by $w$ and get the
equation (124) at once. This leads naturally to the formal form of
the perturbation differential equation. On the other hand, we must
point out that if we do not take the variable transformation, we can
not get directly the formal form by the renormalization (group)
method.

In [30], Kunihiro used the geometrical formula of RG method to get
the reduction equations of dynamical systems. Here we show how to
apply the TR method to study the reduction of the dynamical systems
through a typical example.  Consider the deformed Latka-Volterra
equation
\begin{equation}
\frac{dx}{dt}+by=-\epsilon xy,
\end{equation}
\begin{equation}
\frac{dy}{dt}-ax=\epsilon xy
\end{equation}
which is equivalent to the Lotka-Volterra equation by suitable
variable transformations. In [30], Kunihiro obtained its reduction
equation by RG method. Now we use the TR method to give easily the
result. By substituting $x=x_0+\epsilon x_1+\epsilon^2x_2+\cdots$
and $y=y_0+\epsilon y_1+\epsilon^2y_2+\cdots$ into the above
equations system, we have
\begin{equation}
x_0=W(t_0)e^{i\omega t}+c.c, y_0=-\frac{\omega}{b}(iW(t_0)e^{i\omega
t}+c.c.),
\end{equation}
where $c.c.$ denotes the conjugate terms. In order to find the
equation of amplitude $W$, we only to list the solutions of $x$.
Furthermore, we get the first and second order solutions of $x$ as
follows
\begin{equation}
x_1=\frac{2\omega -ib}{3b\omega}W^2(t_0)e^{i2\omega t}+c.c.,
\end{equation}
and
\begin{equation}
x_2=\frac{-i}{6\omega}(\frac{b^2+\omega^2}{b^2}|W|^2W(t-t_0)e^{i\omega
t}+\frac{W^3}{8b^2\omega^2}(3\omega^2-b^2-4ib\omega)e^{i3\omega
t}+c.c.).
\end{equation}
 Rearranging the solution, we get
\begin{equation*}
x=We^{i\omega t}+\epsilon\frac{2\omega
-ib}{3b\omega}W^2(t_0)e^{i2\omega
t}-\epsilon^2i\frac{\omega^2-b^2-4ib\omega^3}{48b^2\omega^3}W^3e^{i3\omega
t}
\end{equation*}
\begin{equation}
-\epsilon^2i\frac{b^2+\omega^2}{6b^2\omega}|W|^2W(t-t_0)e^{i\omega
t}+c.c.+O(\epsilon^3).
\end{equation}
Expanding every exponent function at the point $t_0$, the
approximate closed renormalization equation is
\begin{equation}
\frac{dW}{dt}=-\epsilon^2i\frac{b^2+\omega^2}{6b^2\omega}|W|^2W,
\end{equation}
which is just the reduction equation of the Lotka-Volterra equation.
Other equations can be dealt with similar method. From the example,
we can see that the reduction equation of a dynamical system will be
obtained from the renormalization equation which is the relation
between the first two terms in Taylor expansion of solution. All
these come from the simple facts in Taylor series.

\section{Homotopy renormalization method and applications}

\subsection{Examples of the weakness of the standard RG method}

The RG method is not omnipotent, and has also some weaknesses. We
list several typical examples to illustrate these weaknesses.

 \textbf{ Example 1}. We consider the equation $y'(t)=\epsilon
(1-y^2(t))$. By the standard renormalization group method, taking
\begin{equation}
y=\sum_{n=0}^{+\infty}y_n\epsilon^n,
\end{equation}
and substituting it into the above equation yields
\begin{equation}
y'_0=0,
\end{equation}
and
\begin{equation}
y'_1=1-y^2_0,
\end{equation}
and so on. From these two equations, we have
\begin{equation}
y_0=A,
\end{equation}
\begin{equation}
y_1=(1-A^2)(t-t_0).
\end{equation}
Therefore,
\begin{equation}
y=A+\epsilon(1-A^2)(t-t_0)+O(\epsilon^2),
\end{equation}
which including a secular term $t-t_0$. By considering $A$ as a
function of $t_0$ and then according to the renormalization group
method,  the RG equation
\begin{equation}
\frac{\partial y}{\partial t_0}=0
\end{equation}
gives
\begin{equation}
A'(t_0)=\epsilon (1-A^2(t_0)).
\end{equation}
This equation is just the same one with the original equation
$y'(t)=\epsilon (1-y^2(t))$. Therefore, this is a cyclic process and
then we can not improve the solution in global.

\textbf{Example 2}. We consider equation $y''(t)=\epsilon
(y^2(t)-y^3(t))$ which has two trivial solutions $y=0$ and $y=1$.
This is a multi-solutions problem.

By the standard RG method, the first two perturbation equations are
\begin{equation}
y''_0=0,
\end{equation}
and
\begin{equation}
y''_1=y^2_0-y^3_0.
\end{equation}
Solving these two equations give
\begin{equation}
y_0=A(t-t_0)+B,
\end{equation}
\begin{equation}
y_1=\frac{1}{12A^2}(A(t-t_0)+B)^4-\frac{1}{20A^2}(A(t-t_0)+B)^5.
\end{equation}
So we have
\begin{equation}
y(t,t_0)=A(t-t_0)+B+\epsilon(\frac{1}{12A^2}(A(t-t_0)+B)^4-\frac{1}{20A^2}(A(t-t_0)+B)^5)+O(\epsilon^2).
\end{equation}
Since we can consider $A$ and $B$ as two independent functions of
variable $t_0$, the normalization group equation gives $A'(t_0)=0$
and $B'(t_0)=0$. This means $A$ and $B$ are two constants.
Therefore, we obtain a global solution
\begin{equation}
y=B,
\end{equation}
which is only a trivial solution. So we can not get a nontrivial
global solution by the RG method.

\textbf{Example 3}. Consider the  equation $y'(t)=1-\epsilon
y^2(t)$.

By the standard RG method, the first two perturbation equations are
\begin{equation}
y'_0=1,
\end{equation}
and
\begin{equation}
y'_1=-y^2_0,
\end{equation}
and so on. From these two equations, we have
\begin{equation}
y_0=t-t_0,
\end{equation}
\begin{equation}
y_1=-\frac{1}{3}(t-t_0)^3+C.
\end{equation}
Therefore,
\begin{equation}
y=t-t_0+\epsilon(-\frac{1}{3}(t-t_0)^3+C)+O(\epsilon^2),
\end{equation}
which including several secular terms. According to the
renormalization group method based on the envelop theory, we know
that the RG equation
\begin{equation}
\frac{\partial y}{\partial t_0}=0
\end{equation}
gives
\begin{equation}
C'(t_0)=\frac{1}{\epsilon},
\end{equation}
whose solution is $C(t_0)=\frac{t_0}{\epsilon}$.  Therefore, the
global solution is
\begin{equation}
y=t+O(\epsilon^2).
\end{equation}

The exact solution of the original equation is
\begin{equation}
y=\frac{A_0\exp(2\sqrt\epsilon t)-1}{A_0\exp(2\sqrt\epsilon t)+1},
\end{equation}
from which we have $\lim_{t\rightarrow \pm\infty}=\pm 1$.  This
means that the solution can not be improved by the renormalization
group method.

Finally,  we consider the equation with a high nonlinearity term:
$y'=y^2+t\sin y+\epsilon y$. By the standard renormalization group
method, taking $y=\sum_{n=0}^{+\infty}y_n\epsilon^n$ and
substituting it into the above equation yields the first
perturbation equation
\begin{equation}
y'_0=y_0^2+t\sin y_0.
\end{equation}
However, we can not give the explicit exact solution of the first
equation, and hence it is not convenient to perform the
renormalization process.

  In order to overcome these weaknesses, we propose an iteration method such that we can eliminate these
  weaknesses. In fact,  what we need is only to have a freedom to
  choose the first approximate solution. The iteration method can
  give us this freedom and then can be applied to more
  equations. In section 5.2, we will give this new iteration method namely
  deformed renormalization group method or a more fashion name as the
  homotopy renormalization  method. Some applications are also
  given. We must point out that homotopy theory has been applied
extensively to nonlinear problems and then some famous methods such
as small parameter method[72-75], homotopy continuous
algorithm[76,77], topology degree theory[75] have been introduced.
Moreover, the homotopy method appeared by some different forms in
literatures. The iteration scheme in the present paper has been
given in Stoker's book[75]. Therefore, my work is just to combine
the homotopy method or equivalent iteration scheme with the
renormalization method together to find the global asymptotic
solutions. Fortunately, this method is efficient and powerful.

\subsection{The homotopy renormalization method}

In the TR and standard RG methods, the first order approximation
$y_0$ is determined by its differential equation, and then it is
fixed. If we can freely choose the first approximation $y_0$, we
will have chance to do more than by the standard RG or TR method. In
fact, we can do it by a topological way.  This is a routine and
classical method belonging to Poincar\'{e} and Stoker and so on. In
fact, the famous Poincar\'{e} 's small parameter method is a kind of
homotopy method(see, for example, [75]) in which a small parameter
is introduced to deform a simple equation to the aim equation.
Stoker modified the Duffing's method to deal with the Duffing's
equation by adding a same term in two sides of the equation to give
an iteration scheme and hence eliminate the secular terms to obtain
the relation between amplitude and frequency. Stoker's method can be
considered as a modern version of small parameter method, which is
the starting point of our homotopy renormalization method.  Assume
that we consider the equation
\begin{equation}
N(t,y,y',\cdots)=0,
\end{equation}
where $N$ is a function. Take a simple linear equation
\begin{equation}
L(y)=0,
\end{equation}
where $L$ is in general a linear operator. Next we take the homotopy
equation as follows
\begin{equation}
(1-\epsilon) L(y)+\epsilon N(t,y,y',\cdots)=0,
\end{equation}
where the homotopy parameter $\epsilon$ satisfying $0\leq \epsilon
\leq 1$. We can see that the homotopy equation changes from the
simple equation $L(y)=0$ to the aim equation $N(t,y,y',\cdots)=0$ as
$\epsilon$ changing from $0$ to $1$. Therefore, we can deform the
solution of the simple equation to that of the aim equation. We
expand the solution of the homotopy equation as a power series of
$\epsilon$
\begin{equation}
y(t,\epsilon)=\sum_{k=0}^{+\infty}y_k(t)\epsilon^k,
\end{equation}
where $y_k(t)'s$ are unknown functions. Substituting the solution
into the homotopy equation and equating the coefficients of the
power of $\epsilon$ to be zeroes, we get the linear equations
\begin{equation}
 L(y_{k+1})=L(y_k)-N(t,y_k,y_k',\cdots),
\end{equation}
for $k=0,1,\cdots$. Solving these linear equations give solutions
$y_0,y_1$ and so on. The above steps are the standard procedures in
this topological method. Now we come into a new step. We expand
every $y_k$ as a power series at a general point $t_0$ and rearrange
the solution and then take $\epsilon=1$ at the end, the solution of
original equation is
\begin{equation}
y(t,t_0)=Y_0(t_0,A,B,\cdots)+Y_1(t_0,A,B,\cdots)(t-t_0)+Y_2(t_0,A,B,\cdots)(t-t_0)^2+O(1),
\end{equation}
where $A$ and $B$ and so on are parameters. From this expression, we
know that the solution is just
\begin{equation}
y(t)=Y_0(t,A,B,\cdots).
\end{equation}
In order to use the renormalization  method, we can consider these
parameters as the functions of $t_0$ and determine them by another
relation, that is, the renormalization equation
\begin{equation}
\frac{\partial Y_0}{\partial t_0}=Y_1.
\end{equation}
Through a suitable choice of the closed equations system, we can
solve out these integral constants such as $A$ and $B$, and then
substituting these parameters into the solution $Y_0$ yields the
asymptotic global solution.

\textbf{Remark 3}. There are two basic methods in numerical
mathematics. One is the elimination method which is from complex to
simple. Another is just the iteration method which is from simple to
complex. By iteration process, we deform the solution of a simple
equation to the solution of a complex equation. The homotopy method
is equivalent to the following iteration method. In fact, we can
rewrite the aim equation as follows
\begin{equation}
Ly=Ly-Ny,
\end{equation}
and take iteration
\begin{equation}
Ly_0=0,
\end{equation}
\begin{equation}
Ly_{n+1}=Ly_n-Ny_n,
\end{equation}
for $n=1,2,\cdots$. Then $y=y_0+y_1+\cdots+y_n+\cdots$ is the
solution of the equation $Ny=0$. It is easy to see that this
iteration scheme is equivalent to the homotopy process by taking
$\epsilon=1$ in the last step. Essentially, the homotopy method
given in here is an iteration method.

If the operator $L$ has an inverse $L^{-1}$, then the iteration
scheme becomes
\begin{equation}
y_{n+1}=y_n-L^{-1}Ny_n,
\end{equation}
where $L^{-1}$ is in general an integral operator if $L$ is an
differential operator.

\textbf{Remark 4}. An important problem is whether the iteration
scheme is convergent. However, we have no a general criterion of
convergence for the HTR method. In usual homotopy  methods such as
small parameter method in quantum field theory and in differential
equations, a routine trick is to take analytic continuation of the
small parameter. In the HTR method, this is also a possible choice
of getting the resulting solutions. For the concrete applications,
since in general the asymptotic solutions include the introduced
small parameter by the analytic forms, we can obtain the resulting
solutions by direct taking the parameter to be one.

\subsection{Applications of HTR method}

\textbf{Example 1}. Consider the simple multi-solutions problem
$y'(t)=1- y^2(t)$.

Its trivial global solutions are $y=\pm 1$ which can be obtained by
the standard RG method. Now we consider the  terminal value
condition $y(+\infty)=1$.  For the purpose of finding the
non-trivial asymptotic solution, we use the HTR method. The homotopy
equation is give by
\begin{equation}
y'+y=1-\epsilon(y^2-y).
\end{equation}
Taking
\begin{equation}
y=\sum_{n=0}^{+\infty}y_n\epsilon^n,
\end{equation}
and substituting it into the above homotopy equation yields
\begin{equation}
y'_0+y_0=1,
\end{equation}
and
\begin{equation}
y'_1+y_1=y_0-y^2_0,
\end{equation}
and so on. From these two equations, we have
\begin{equation}
y_0=A\exp(-t)+1,
\end{equation}
\begin{equation}
y_1=-A(t-t_0)\exp(-t)+A^2\exp(-2t).
\end{equation}
Therefore,
\begin{equation}
y=A\exp(-t)+1+\epsilon(-A(t-t_0)\exp(-t)+A^2\exp(-2t))+O(\epsilon^2).
\end{equation}
 According to the TR method, we know that the
renormalization equation
\begin{equation}
\frac{\partial}{\partial
t_0}(A\exp(-t_0)+1+A^2\epsilon\exp(-2t_0))=-A\epsilon\exp(-t_0)-A\exp(-t_0)-2A^2\epsilon\exp(-2t_0)
\end{equation}
gives
\begin{equation}
A'(t_0)+\epsilon A=0,
\end{equation}
whose solution is $A(t_0)=A_0\exp(-\epsilon t_0)$.  Therefore, the
global solution is
\begin{equation}
y=1+A_0\exp(-(1+\epsilon)t)+\epsilon
A_0^2\exp(-2(1+\epsilon)t))+O(\epsilon^2).
\end{equation}
By taking $\epsilon=1$, we have
\begin{equation}
y=1+A_0\exp(-2t)+A_0^2\exp(-4t))+O(1),
\end{equation}
which is a non-trivial global solution.

For the  terminal value condition, the exact solution is
\begin{equation}
y=\frac{1-A\exp(-2t)}{1+A\exp(-2t)}=1+2A\exp(-2t)+2A^2\exp(-4t)+\cdots.
\end{equation}
Comparing the asymptotic solution with the exact solution, when $t$
tends to infinity,  these two solutions are the same one. This shows
that the HTR method is efficient for this problem.

\textbf{Example 2}. Forced Duffing equation[72] $y''(t)=-\alpha
y+\beta y^3(t)+F\cos(\omega t)$.

By the HTR method, we take homotopy equation
\begin{equation}
y''+\omega^2 y=\epsilon((\omega^2-\alpha)y+\beta y^3+F\cos(\omega
t)),
\end{equation}
and the solution
\begin{equation}
y=\sum_{n=0}^{+\infty}y_n\epsilon^n.
\end{equation}
Substituting it into the above homotopy equation yields
\begin{equation}
y''_0+\omega^2 y_0=0,
\end{equation}
and
\begin{equation}
y''_1+\omega^2 y_1= (\omega^2-\alpha)y_0+\beta y_0^3+F\cos(\omega
t),
\end{equation}
and so on. From these two equations, we have
\begin{equation}
y_0=A\cos(\omega t+\theta),
\end{equation}
\begin{equation}
y_1=(\frac{\omega^2-\alpha}{2\omega}A+\frac{3\beta}{8\omega}A^3)(t-t_0)\sin(\omega
t+\theta)-\frac{\beta}{36\omega^2}A^3\cos(3(\omega
t+\theta))+\frac{F}{2\omega}(t-t_0)\sin(\omega t).
\end{equation}
Therefore,
\begin{equation*}
y=A\cos(\omega
t+\theta)+\epsilon((\frac{\omega^2-\alpha}{2\omega}A+\frac{3\beta}{8\omega}A^3)(t-t_0)\sin(\omega
t+\theta)
\end{equation*}
\begin{equation}
-\frac{\beta}{36\omega^2}A^3\cos(3(\omega
t+\theta))+\frac{F}{2\omega}(t-t_0)\sin(\omega t))+O(\epsilon^2).
\end{equation}
By taking $\epsilon=1$, the renormalization equation is given by
\begin{equation*}
\frac{\partial}{\partial t_0}(A\cos(\omega
t_0+\theta)-\frac{\beta}{36\omega^2}A^3\cos(3(\omega t_0+\theta)))
\end{equation*}
\begin{equation}
=(\frac{\omega^2-\alpha}{2\omega}A+\frac{3\beta}{8\omega}A^3)\sin(\omega
t_0+\theta)+\frac{F}{2\omega}\sin(\omega t_0)-A\sin(\omega
t_0+\theta)+\frac{\beta}{12\omega^2}A^3\sin(3(\omega t_0+\theta))
\end{equation}
from which a closed equations system can be taken as
\begin{equation}
A'(t_0)=0, \theta'(t_0)=0,
\end{equation}
with $A$ and $\theta$ satisfying the following algebraic equations
\begin{equation}
\theta(t_0)=0,
\end{equation}
\begin{equation}
\frac{\omega^2-\alpha}{2\omega}A+\frac{3\beta}{8\omega}A^3+\frac{F}{2\omega}=0.
\end{equation}
Under this condition, a global solution is
\begin{equation}
y=A\cos(\omega t)-\frac{\beta}{36\omega^2}A^3\cos(3(\omega t))+O(1),
\end{equation}
where $A_0$ is the real root of the equation
$4(\omega^2-\alpha)A+3\beta A^3+4F=0$.

This solution is same with the Stoker's result in [75]. This means
the HTR method is efficient for the problem.

\textbf{Example 3}. $y''(t)=\eta(y^3(t)- y^2(t))$.

This is a multi-solutions problem. Two trivial solutions are $y=0$
and $y=1$. By the standard RG method, we only give these two trivial
solutions. In order to give non-trivial global solution, we use the
HTR method. We will give two choices of the homotopy equation. The
first homotopy equation is given by
\begin{equation}
y''+y=\epsilon(y+\eta(y^3-y^2)).
\end{equation}
Then taking
\begin{equation}
y=\sum_{n=0}^{+\infty}y_n\epsilon^n,
\end{equation}
and substituting it into the above homotopy equation yields
\begin{equation}
y''_0+y_0=0,
\end{equation}
and
\begin{equation}
y''_1+y_1=y_0+\eta(y_0^3-y^2_0),
\end{equation}
and so on. From these two equations, we have
\begin{equation}
y_0=A\cos(t+\theta),
\end{equation}
\begin{equation}
y_1=\frac{1+\eta}{2}A(t-t_0)\sin(t+\theta)+\frac{\eta}{6}A^2\cos(2(t+\theta))+\frac{\eta}{2}A^2.
\end{equation}
Therefore,
\begin{equation}
y=A\cos(t+\theta)+\epsilon(\frac{1+\eta}{2}A(t-t_0)\sin(t+\theta)+\frac{\eta}{6}A^2\cos(2(t+\theta))
+\frac{\eta}{2}A^2)+O(\epsilon^2),
\end{equation}
We know that the renormalization equation
\begin{equation*}
\frac{\partial}{\partial
t_0}(A\cos(t_0+\theta)+\frac{\eta}{6}\epsilon A^2\cos(2(t_0+\theta))
+\frac{\eta}{2}\epsilon A^2)
\end{equation*}
\begin{equation}
=-A\sin(t_0+\theta)-\frac{\eta}{3}\epsilon
A^2\sin(2(t+\theta))+\epsilon\frac{1+\eta}{2}A\sin(t+\theta)
\end{equation}
gives a closed equations system
\begin{equation}
A'(t_0)=0,
\end{equation}
\begin{equation}
\theta'(t_0)+\frac{1+\eta}{2}=0,
\end{equation}
whose solutions are $A(t_0)=A_0$ and $\theta=-\frac{1+\eta}{2}t_0$.
Therefore, the global solution is
\begin{equation}
y=A_0\cos(\frac{1-\eta}{2}t)+\epsilon(\frac{\eta}{6}A_0^2\cos(\frac{1-\eta}{2}t)+\frac{\eta}{2}A_0^2)+O(\epsilon^2),
\end{equation}
By taking $\epsilon=1$, we have
\begin{equation}
y=(A_0+\frac{\eta}{6}A_0^2)\cos(\frac{1-\eta}{2}t)+\frac{\eta}{2}A_0^2+O(1),
\end{equation}
which is just a non-trivial global solution. This is also a periodic
solution.

Next we take the second homotopy equation to find its asymptotic
solution. Firstly we take a variable transformation
\begin{equation}
y=z+\frac{1}{3},
\end{equation}
and the equation becomes
\begin{equation}
z''=\eta(z^3-z-\frac{2}{27}).
\end{equation}
The homotopy equation is given by
\begin{equation}
z''+\omega^2z=\epsilon(\omega^2z+\eta(z^3-z-\frac{2}{27})).
\end{equation}
By the similar method with the above example 2, we have
\begin{equation*}
z=A\cos(\omega
t+\theta)+\epsilon((\frac{\omega^2-\eta}{2\omega}A+\frac{3\eta}{8\omega}A^3)(t-t_0)\sin(\omega
t+\theta)
\end{equation*}
\begin{equation}
-\frac{\beta}{36\omega^2}A^3\cos(3(\omega
t+\theta))-\frac{2}{27\omega^2}+O(\epsilon^2).
\end{equation}
By taking $\epsilon=1$. the renormalization equation is given by
\begin{equation*}
\frac{\partial}{\partial t_0}(A\cos(\omega
t_0+\theta)-\frac{\eta}{36\omega^2}A^3\cos(3(\omega t_0+\theta)))
\end{equation*}
\begin{equation}
=(\frac{\omega^2-\eta}{2\omega}A+\frac{3\eta}{8\omega}A^3)\sin(\omega
t_0+\theta)-A\sin(\omega
t_0+\theta)+\frac{\eta}{12\omega^2}A^3\sin(3(\omega t_0+\theta)),
\end{equation}
from which a closed equations system can be taken as
\begin{equation}
A'(t_0)=0, \theta'(t_0)=0,
\end{equation}
with $A$ and $\theta$ satisfying the following algebraic equations
\begin{equation}
\theta(t_0)=0,
\end{equation}
\begin{equation}
\frac{\omega^2-\alpha}{2\omega}A+\frac{3\eta}{8\omega}A^3=0.
\end{equation}
Under this condition, a global solution is
\begin{equation}
y=A_0\cos(\omega t)-\frac{\eta}{36\omega^2}A_0^3\cos(3\omega
t)-\frac{2}{27\omega^2}+O(1),
\end{equation}
where $A_0$ is the real root of the equation
$\omega^2=\eta-\frac{3\eta}{4}A^2$. This means that the asymptotic
solution of $y$ is also a periodic solution.

By complete discrimination system for polynomial and integral
method[79], we can give all solutions of the ordinary differential
equation in the example, among these solutions, there are elliptic
function solutions and a hyperbolic function solution. Therefore,
our asymptotic solutions can be considered as the approximations to
elliptic solutions which are double periodic functions.

\textbf{ Example 4}. We consider the famous Blasius
equation[57,59,60, 80] $y'''+yy''=0,y(0)=y'(0)=0,y'(+\infty)=1/2$.

By the HTR method, take homotopy equation
\begin{equation}
y'''+ y''=\epsilon(y''-yy''),
\end{equation}
and the solution
\begin{equation}
y=\sum_{n=0}^{+\infty}y_n\epsilon^n.
\end{equation}
Substituting it into the above homotopy equation yields
\begin{equation}
y'''_0+ y''_0=0,
\end{equation}
and
\begin{equation}
y'''_1+ y''_1= y''_0-y_0y''_0,
\end{equation}
and so on. From these two equations, we have
\begin{equation}
y_0=A\mathrm{e}^{-t}+B(t-t_0)+C,
\end{equation}
\begin{equation}
y_1=A(2-2C-3B)\mathrm{e}^{-t}+\frac{A^2}{4}\mathrm{e}^{-2t}+A(1-C-2B)(t-t_0)\mathrm{e}^{-t}-
\frac{AB}{2}(t-t_0)^2\mathrm{e}^{-2t}.
\end{equation}
Therefore,
\begin{equation*}
y=A\mathrm{e}^{-t}+B(t-t_0)+C+\epsilon(A(2-2C-3B)\mathrm{e}^{-t}+\frac{A^2}{4}\mathrm{e}
^{-2t}
\end{equation*}
\begin{equation}
+A(1-C-2B)(t-t_0)\mathrm{e}^{-t}-
\frac{AB}{2}(t-t_0)^2\mathrm{e}^{-2t})+O(\epsilon^2).
\end{equation}
The renormalization equation is
\begin{equation*}
\frac{\partial }{\partial
t_0}\{A\mathrm{e}^{-t_0}+C+\epsilon(A(2-2C-3B)\mathrm{e}^{-t_0}+\frac{A^2}{4}\mathrm{e}
^{-2t_0})\}
\end{equation*}
\begin{equation}
=A(1-C-2B)\epsilon\mathrm{e}^{-t_0}-A\mathrm{e}^{-t_0}-\epsilon(A(2-2C-3B)\mathrm{e}^{-t_0}-\frac{A^2}{2}\mathrm{e}
^{-2t_0}).
\end{equation}
A closed equations system can be taken as
\begin{equation}
B'(t_0)=0, C'(t_0)=0,
\end{equation}
\begin{equation}
A'(t_0)=\frac{\epsilon(1-C-2B)}{1+\epsilon(2-2C-3B)}A,
\end{equation}
whose solutions are
\begin{equation}
B=B_0, C=C_0,
\end{equation}
\begin{equation}
A(t_0)=A_0\exp(\frac{\epsilon(1-C-2B)}{1+\epsilon(2-2C-3B)}t_0).
\end{equation}
Therefore, by taking $\epsilon=1$, the global solution is
\begin{equation}
y=A_0(3-2C_0-3B_0)\exp(\frac{C_0+B_0-2}{3-2C_0-3B_0)}t)+\frac{A_0^2}{4}\exp(\frac{2(C_0+B_0-2)}{3-2C_0-3B_0)}t)+C_0,
\end{equation}
which is an approximate but global solution. Furthermore, if we
assume condition $y'(+\infty)=\frac{1}{2}$ which is the real
condition for Blasius equation, we can not direct use the above
solution. We need direct consider the derivative of the solution
$y'(t)$. From the solution (226), we have
\begin{equation*}
y'=-A\mathrm{e}^{-t}+B+\epsilon(A(B+C-1)\mathrm{e}^{-t}-\frac{A^2}{2}\mathrm{e}
^{-2t}
\end{equation*}
\begin{equation}
-A(1-C-2B)(t-t_0)\mathrm{e}^{-t}-
AB(t-t_0)\mathrm{e}^{-2t}+AB(t-t_0)^2\mathrm{e}^{-2t})+O(\epsilon^2).
\end{equation}
By the similar computation, from the renormalization equation, a
closed equations system can be taken as
\begin{equation}
B'(t_0)=0, C'(t_0)=0,
\end{equation}
\begin{equation}
A'(t_0)(B+C-2)+A(1-C-2B)=0,
\end{equation}
where we have taken $\epsilon=1$. Solving these equations give
\begin{equation}
B=B_0, C=C_0,
\end{equation}
\begin{equation}
A(t_0)=A_0\exp(\frac{1-C_0-2B_0}{B_0+C_0-2}t_0),
\end{equation}
\begin{equation}
y'=A_0(B_0+C_0-2)\exp(\frac{3-2C_0-3B_0)}{C_0+B_0-2}t)-\frac{A_0^2}{2}\exp(\frac{2(3-2C_0-3B_0)}{C_0+B_0-2}t)+C_0+B_0.
\end{equation}
According to $y'(+\infty)=\frac{1}{2}$, we have
$B_0+C_0=\frac{1}{2}$. Furthermore, by taking $C_0=0$,  the
corresponding solution is
\begin{equation}
y'=-\frac{3}{2}A_0\exp(-t)-\frac{A_0^2}{2}\exp(-2t)+\frac{1}{2}.
\end{equation}
Therefore, by integrating it once, the global approximate solution
is given by
\begin{equation}
y=\frac{3}{2}A_0\exp(-t)+\frac{A_0^2}{4}\exp(-2t)+\frac{1}{2}t.
\end{equation}
Although this solution includes a secular term $\frac{1}{2}t$, but
this is the need of the real physical problem such that it can
satisfy the boundary condition at infinity. So we should use the
renormalization  method to adapt new cases in practice. Comparing
our result with the solutions obtained by HAM method and general
series method, we find these solutions have the same asymptotic
behaviors. Therefore, the HTR method is efficient for the problem.

\textbf{Remark 5}.  By the above examples, we can see that the HTR
method can be
 applied to many equations. In final, we give a discussion on the equation (156). We can take the homotopy equation as
 \begin{equation}
y'+y=\eta(y+y^2+t\sin y+\epsilon y).
 \end{equation}
Corresponding first and second equation are given by
\begin{equation}
y_0'+y_0=0,
 \end{equation}
 \begin{equation}
y_1'+y_1=\eta(y_0+y_0^2+t\sin y_0+\epsilon y_0).
 \end{equation}
 By substituting the solution $y_0=\exp(-(t-t_0))A(t_0)$ into the
 second equation to get $y_1$. Furthermore, expanding $\sin y_0$ by
 a power series and using the TR method, we can give the asymptotic
 solution. Here we omit them.

\section{ Conclusions}

The TR method is based on Taylor expansion which has a strict
mathematical foundation, and it recovers the essence of the standard
RG method. Since the secular terms are automatically eliminated, the
TR method shows its advantages from both of theory and applications.
Although the TR method give the same result with the RG method in
practice,
 but the TR method
is more simple and direct than the RG method, and seems to be more
efficient and accurate in practice than other methods in extracting
global information series by eliminating the secular terms.
Moreover, the most important point is that logic of the TR method is
more clear than the RG method and its geometrical formula. However,
the TR method (or RG method) is not omnipotent, and also has some
weaknesses. In order to overcome these weaknesses, a modified TR
method namely homotopy renormalization method is proposed.
Furthermore, the homotopy renormalization method offer a freedom of
choosing the first approximate solution in perturbation expansion so
that we can improve the global approximate solutions for nonlinear
differential equations. Even the considered equations do not include
a small parameter, we can yet use this method to get the global
asymptotic solutions. The TR and HTR methods show that behind the RG
method exists a very simple and clear mathematical construction that
is just the Taylor expansion so that these methods can be applied
easily to many perturbation problems.

\textbf{Acknowledgements}. Thanks to anonymous referees for their
valuable comments on the first version of the paper. I would like to
thank Prof. Nayfeh for his helpful suggestions. I also appreciate the Editor's kindly help.

\end{document}